\documentclass{aa}  
\usepackage{graphicx}
\usepackage{xcolor}
\usepackage{natbib}
\bibpunct{(}{)}{;}{a}{}{,}
\usepackage{txfonts}
\usepackage[citecolor=blue, linkcolor=blue, urlcolor = blue, colorlinks = true]{hyperref}
\usepackage{listings}
\usepackage{lscape}
\usepackage{minted}
\usepackage{placeins}
\usepackage{amsmath}
\usepackage{xspace}
\usepackage{subfigure}
\usepackage{cancel}
\usepackage{todonotes}
\newcommand{\incom}[1]{}
\usepackage{xcolor}
\hypersetup{
    colorlinks=true,
    linkcolor=blue,
    citecolor=blue,
    filecolor=magenta,
    urlcolor=cyan,
    }

\newcommand{\code}[1]{\texttt{#1}}

\newcommand{\acor}{\code{ACOR}\xspace}

\begin{document}

   \title{Probing magnetic fields in stars: \acor, an oscillation framework including rotation and magnetism}

   \author{A. Fort
          \inst{1}
          \and
         R.-M. Ouazzani\inst{1}
         \and
         L. Barrault\inst{2}
         \and
         L. Manchon\inst{1}
         \and
         L. Petitdemange\inst{1}
          }

   \institute{LIRA, Observatoire de Paris,Université Paris Cité, Université PSL, CNRS, Sorbonne Université, CY Cergy Paris Université, 92190 Meudon, France\\
              \email{antoine.fort@obspm.fr}
        \and 
        Institute of Science and Technology Austria (ISTA), Am Campus 1, 3400 Klosterneuburg, Austria
        }

   \date{Received Month date, year; accepted month date, year}

  \abstract
   {
   Understanding the role of internal magnetic fields in stars remains a major challenge for the description of angular momentum transport and stellar evolution. It is therefore essential to probe these magnetic fields within the star, and asteroseismology provides a powerful means to do so.
   }
   {
In this work, we present a new implementation of the 2D oscillation code \acor (\textit{Adaptive Code of Oscillations towards Realistic modeling}), which incorporates the effects of both stellar rotation and magnetic fields.
   }
   {
The code has been rendered modular, making it possible to specify the set of equations to solve and the assumptions through a symbolic calculus approach. The full set of adiabatic, non-radial pulsation equations is solved using a spectral approach for the angular part of the modes and high-order finite differences for the radial part. As a first step, we focus on a magnetic field that is purely toroidal and axisymmetric about the star's rotation axis. The numerical results are compared against first-order perturbative predictions in the weak-field regime and with the traditional approximation of rotation and magnetism (TARM) in the case of stronger magnetic fields.
   }
   {
We validate this new implementation of \acor against perturbative and TARM approaches, showing good agreement and demonstrating the robustness of the code. The breakdown of these methods provides general validity limits. We show that internal magnetic fields leave signatures in the period spacing of g modes. These features provide a promising seismic diagnostic to probe deep stellar magnetism in $\gamma$ Dor stars. Future work will aim to extend this framework to more realistic magnetic field topologies and a broader range of pulsating stars, including red giants.
   }
   {}


   \maketitle
   \titlerunning{ACOR: stellar oscillations with rotation and magnetism}
   \authorrunning{Fort et al.}
   \nolinenumbers

\defcitealias{ouazzani_pulsations_2012}{O12}
\defcitealias{ouazzani_pulsations_2015}{O15}
\defcitealias{dhouib_detecting_2022}{D22}

\section{Introduction}

Rotation plays a fundamental role in stellar evolution through a wide range of processes, including centrifugal deformation, rotational mixing, internal flows, angular momentum transport, and magnetic field generation among others. Accurate modeling of it plays a central role in our understanding of stellar evolution.
Until about 15 years ago, rotation was poorly constrained by observations. This has changed thanks to asteroseismology and the {\it Kepler} mission \citep{borucki_kepler_2010}, which provided information on the internal rotation of thousands of stars. Seismic measurements revealed that angular momentum transport in stars is significantly more efficient than predicted by purely hydrodynamic models \citep{eggenberger_angular_2012,eggenberger_asteroseismology_2019,ceillier_understanding_2013,Marques2013,ouazzani__2019}. For instance, these models fail to reproduce the slow core rotation observed in stars on the red giant branch (RGB) \citep{beck_fast_2012,deheuvels_seismic_2012,mosser_probing_2012,cantiello_2014_angular_momentum} and on the main sequence (MS) \citep{christophe_deciphering_2018,ouazzani__2019,li_period_2019,li_gravity-mode_2020}. 
This discrepancy has prompted the community to explore additional mechanisms capable of efficiently extracting angular momentum from stellar cores such as wave-driven transport and transport by oscillation modes \citep{belkacem_angular2_2015,belkacem_angular_2015,rogers_differential_2015,pincon_can_2017,aerts_AMtransport_2019,bordadagua_efficiency_2025}. Processes induced by internal magnetic fields are among the serious contenders to explain such angular momentum transport \citep{rudiger_angular_2015,jouve_three-dimensional_2015,fuller_slowing_2019,takahashi_modeling_2021,petitdemange_spin-down_2023,petitdemange_taylerspruit_2024,barrere_tayler-spruit_2025}. In this context, placing observational constraints on the strength and geometry of internal magnetic fields has become an essential step toward understanding angular momentum transport in stars.

Spectropolarimetry provides direct measurements of stellar surface magnetic fields by measuring Zeeman-induced polarization signatures \citep[see for instance][and references therein]{neiner_origin_2015}. However, this method remains insensitive to magnetic fields within stellar interiors, which are thought to play a critical role in angular momentum transport inside radiative zones.

Asteroseismology has proven essential in providing constraints on internal magnetic fields. The first indirect detection of magnetic fields based on oscillations was reported for a subset of red giant stars from the \textit{Kepler} mission. These stars display in their pulsation spectra dipolar modes with significantly reduced amplitudes, while their radial modes remain unaffected \citep{mosser_characterization_2012}. Many studies have shown that this damping may result from the presence of strong confined magnetic fields \citep{fuller_asteroseismology_2015,lecoanet_conversion_2017,rui_gravity_2023,muller_oscillations_2025}. However, this interpretation remains debated, particularly due to the observation of stars presenting partially suppressed modes (so-called depressed modes) or unaffected quadrupole mixed modes \citep{mosser_dipole_2017}, which challenge the magnetic damping scenarios.

Regarding the direct effects on the frequencies, the first-order perturbative approach has been studied theoretically \citep{unno_nonradial_1989, gough_effect_1990, hasan_probing_2005} and refined later to take into account various magnetic configurations and the interplay with rotation \citep{gomes_core_2020, bugnet_magnetic_2021, mathis_probing_2021,loi_topology_2021}. This approach has been further developed to relate magnetic signatures on oscillation spectra to the magnetic field topology and strength, particularly in the case of red giants whose slow rotation can be treated perturbatively. Several studies rely on observed asymmetries in rotational multiplets \citep{li_magnetic_2022,li_internal_2023,hatt_asteroseismic_2024}. Another series of studies use magnetically induced frequency deviations from the regular period spacing of g modes to infer the strength of internal magnetic fields \citep{loi_topology_2021,li_magnetic_2022,bugnet_magnetic_2022,deheuvels_strong_2023}. These perturbative analyses have led to the first quantitative estimates of buried magnetic field strengths, ranging from tens to hundreds of kilogauss, near the hydrogen-burning shell \citep{deheuvels_strong_2023,bhattacharya_detectability_2024}. Nonetheless, seismic detections remain challenging due to degeneracies between magnetic and rotational signatures \citep{mathis_asymmetries_2023} and the complexity of field topologies \citep{das_unveiling_2024}, which are often simplified to allow tractable perturbative solutions.

While red giants have been the primary focus for seismic magnetic diagnostics, other classes of pulsating stars offer promising opportunities. One such class is MS $\gamma$ Dor stars, which have masses ranging from $1.3M_\odot$ to $2.0M_\odot$ depending on their evolutionary stage, and a radiative zone just above the convective core. They are direct progenitors of RGB stars, making it possible to probe internal magnetic fields through stellar evolution. $\gamma$ Dor stars exhibit both gravito-inertial and Rossby modes driven by the convective blocking mechanism \citep{dupret_convection-pulsation_2005} and probe the radiative zone just above the convective core where magnetic fields are likely to emerge from convective dynamos. These modes have been clearly detected in over 600 $\gamma$ Dor stars observed over four years by {\it Kepler} \citep{li_gravity-mode_2020} and in about 60 observed by TESS \citep{garcia_internal_2022}, and they present low frequencies where magnetic effects on oscillations are expected to be most significant. However, due to their fast rotation, the approaches used for red giants are not directly applicable here. This underlines the need for a theoretical framework capable of describing the influence of arbitrary internal magnetic fields in rapidly rotating stars, to enable the detection and interpretation of magnetic signatures in $\gamma$ Dor pulsation spectra.

As a result, current efforts are increasingly directed toward developing non-perturbative frameworks capable of capturing the full influence of strong magnetic fields and rotation on stellar oscillations. For moderate rotation, the classical approach is the traditional approximation of rotation (TAR) (in geophysics \citet{c_eckart_hydrodynamics_1960}; later applied for stellar pulsations in \citet{berthomieu_78}). It simplifies the problem by neglecting the horizontal component of the rotation vector, which is justified in radiative zones due to the strong radial stratification. This allows the Coriolis force to be included, while keeping the eigenvalue problem separable in spherical coordinates.
The TAR has proven to be a powerful tool for interpreting the oscillation spectra of $\gamma$ Dor stars, enabling mode identification and the determination of internal rotation rates \citep{bouabid_effects_2013,reeth_gravity-mode_2015,reeth_interior_2016,christophe_deciphering_2018,li_period_2019,li_gravity-mode_2020}. However, the TAR relies on simplifying assumptions that limit its applicability, particularly in the case of rapid rotation with strong centrifugal distortion \citep{ballot_2d_2012,ouazzani_new_2017}, or in convectively unstable regions \citep{ouazzani_first_2020}. To overcome these limitations, several studies have adopted approaches that incorporate the complete effects of rotation on stellar oscillations (\citealp{reese_acoustic_2006,reese_pulsation_2009,ouazzani_pulsations_2012,ouazzani_pulsations_2015}, hereafter \citetalias{ouazzani_pulsations_2012} and \citetalias{ouazzani_pulsations_2015}).

In the context of the magnetic effects on gravito-inertial modes, several studies have explored how magnetic fields modify the predictions of the TAR. Some of these works consider rotation through the TAR framework, while incorporating the effects of magnetic field as a perturbative correction. Both aligned and inclined magnetic field configurations relative to the rotation axis have been investigated within this approach \citep{prat_period_2019, prat_effect_2020, lignieres_perturbative_2024}. Building on this, recent efforts have extended the theoretical framework to propose a magnetically modified version of the TAR (hereafter TARM), which aims to better capture the combined influence of rotation and magnetism on g-mode propagation in stellar interiors (\citealp{mathis_2011_TARM};\citealp{dhouib_detecting_2022} hereafter \citetalias{dhouib_detecting_2022};\citealp{rui_asteroseismic_2024}). Going beyond the TARM approximation would allow us to assess its validity, explore magnetic regimes, and probe core magnetism through mode coupling \citep{barrault_exploring_2025}.

In this work, we build upon these efforts by extending the 2D non-perturbative stellar oscillation code \acor (\citetalias{ouazzani_pulsations_2012},\citetalias{ouazzani_pulsations_2015}) to include the influence of internal magnetic fields. Because the code now includes more than just the effects of rotation, we redefine its acronym: \acor now stands for {\it Adaptive Code of Oscillations towards Realistic modeling}. In the new version presented here, specifically, we implement the Lorentz force within the framework of ideal magnetohydrodynamics (MHD) assuming a confined, toroidal magnetic field that is axisymmetric about the rotation axis, thereby providing a more comprehensive framework for studying magneto-rotational effects on stellar oscillations. Here, “confined” refers to a magnetic field that is weak in the outer layers, with a near-zero surface strength, such that its impact on the stellar structure is negligible. This choice of a confined magnetic field is relevant in the case of $\gamma$ Dor because no surface magnetic field has been detected with spectropolarimetry above a threshold of $100$~G \citep{thomson-paressant_search_2023,labadie-bartz_results_2025}.

The origin of these magnetic fields may be attributed either to fossil fields inherited from earlier evolutionary phases \citep[][and references therein]{braithwaite_review_17}, or to dynamo processes \citep[][and references therein]{kapyla_2025} sustained by internal fluid motions such as differential rotation and convective turbulence \citep[see][and the references therein in the context of asteroseismology of $\gamma$ Dor pulsating stars]{barrault_exploring_2025,takata_asteroseismic_2025}. In radiative regions of intermediate-mass stars, both scenarios are viable and may coexist, leading to complex magnetic configurations shaped by the interplay between rotation, stratification, and magnetic instabilities. All these scenarios involve a toroidal component, which, under certain conditions, can become particularly strong. For instance, in the Tayler–Spruit dynamo mechanism \citep{tayler_adiabatic_1973,spruit_differential_1999,spruit_dynamo_2002}, differential rotation is able to wind up an initial poloidal magnetic field to a toroidal one. It creates conditions prone to the Tayler instability that will regenerate the poloidal magnetic field. Recent simulations of this process \citep{petitdemange_spin-down_2023,petitdemange_taylerspruit_2024, barrere_tayler-spruit_2025} demonstrate the formation of a strong, confined toroidal field. This choice to explore purely toroidal magnetic fields as a first step also offers significant numerical advantages.

This paper aims to present a model that accurately captures the combined effects of rotation and magnetic fields on stellar oscillations. Section \ref{sec:analytical} presents the pulsation equations in spheroidal geometry and the magnetic field models used in this study. Section \ref{sec:numerical_method} introduces the numerical method specific to the extension of the \acor code and presents convergence tests.  Section \ref{sec:perturbative} showcases the application of the code to a $\gamma$ Dor star model in slow rotation, comparing these results with those from first-order perturbative theory. Section \ref{sec:dhouib} presents a comparison with the work of \citetalias{dhouib_detecting_2022} considering a TARM approach. Section \ref{sec:discussions} discusses the limits of the model implemented here, and then presents some comparisons between the results obtained here and the current state of the measurements. Finally, Section \ref{sec:conclusion} summarizes the main results and provides perspectives for future work. 

\section{Equations in spheroidal geometry}
\label{sec:analytical}

In this section, we present the ideal MHD equations used in this new implementation of \acor for the case of an axisymmetric magnetic field about the rotation axis. The equations are formulated in spheroidal geometry to account for rotational deformation. Details of the original \acor formulation can be found in \citetalias{ouazzani_pulsations_2012} and \citetalias{ouazzani_pulsations_2015}.

\subsection{Spheroidal geometry}
We used the same framework developed in \acor, in which the centrifugal force distorts the stellar structure from spherical to spheroidal geometry. We adopted a multidomain method following \citet{bonazzola_numerical_1998} and \citetalias{ouazzani_pulsations_2015}, with each domain boundary aligned with key structural transitions in the model (e.g., the convective-radiative interface or the stellar surface). In this system, to better adapt to the star's spheroidal geometry, the radial coordinate, $\zeta$, replaces the usual radial distance, $r$. Detailed explanations are available in \citetalias{ouazzani_pulsations_2015}. The same vector basis $(\vec{a}_\zeta,\vec{a}_\theta,\vec{a}_\varphi)$ defined in \citetalias{ouazzani_pulsations_2012} is used, which, in the 1D case, corresponds to the classical spherical vector basis.

As a first step, we assumed that the magnetic field does not significantly deform the stellar structure. This is particularly justified for g modes that are mainly present in deep radiative regions where the gas pressure dominates over the magnetic pressure, making the Lorentz force negligible in the macroscopic force balance \citep{fuller_2023}. In such conditions, the magnetic field affects oscillations directly rather than through modifying the equilibrium structure. Another motivation for this choice is the lack of stellar evolution models suitable as inputs for oscillation codes that consistently incorporate magnetic effects alongside centrifugal distortion. To satisfy this assumption, we consider only confined magnetic fields in this study. This restriction does not reflect a limitation of \acor, which has been designed to treat general 2D axisymmetric structures.

\subsection{Equations}
\label{subsec:sys_eq}
We modeled the stellar interior as an inviscid, self-gravitating fluid, neglecting relativistic effects and binary interactions. To compute the pulsations of rotating magnetic stars, we solved the system of ideal MHD equations. This system includes the mass and energy conservation equations, the momentum equation, the induction equation, and Poisson's equation.  Ideal MHD is appropriate in the deep layers of stars because both the magnetic diffusivity and the kinematic viscosity are small, and nonideal effects occur on timescales much longer than those relevant for stellar oscillations, making it both a physically valid and computationally efficient approximation for asteroseismic modeling. 
We assumed adiabatic oscillations, as thermal processes generally occur on much longer timescales than dynamical ones in most of the stellar interior. This makes heat exchange negligible during pulsations. Although non-adiabatic effects may become important in outer layers of low-mass stars,  or regions where mode excitation occurs, the adiabatic approximation remains valid for studying the star's internal dynamical behaviour.

We assumed that both rotation and magnetic field depend only on the radial coordinate, $\zeta$, and the colatitude, $\theta$. The rotation vector was defined as follows, with $\Omega$ the angular velocity:
\begin{equation}
    \vec{\Omega}=\Omega(\zeta,\theta)(\cos\theta\vec{a_\zeta}-\sin\theta\vec{a_{\theta}})
.\end{equation}
The equilibrium velocity field, written $\vec{v_0}$, was assumed to result solely from rotation and defined as
\begin{equation}
    \vec{v_0}=\vec{\Omega}\times\vec{r}
.\end{equation}
This assumption is physically justified in radiative zones of intermediate- and high-mass stars, where the average flow is largely stable and dominated by rotation. While internal magnetic fields and angular momentum transport processes can in principle induce additional motions, their characteristic timescales are typically much longer than those of oscillations.

We adopted the Eulerian formalism to describe small perturbations and, following the approach of \cite{unno_nonradial_1989}, linearized the MHD equations around a stationary equilibrium state. The system of equations to be solved (in CGS units) was as follows: 
\begin{equation}
    \partial_t\rho'+\vec{\nabla}\cdot\left(\rho_0 \vec{v'}+\rho' \vec{v_0}\right)=0
,\end{equation}
\begin{equation}
    \begin{aligned}
        \left[\left(\partial_t+\Omega\partial_\varphi\right)v_i'\right]{\vec{a}_i}+2\vec\Omega\times\vec{v'}+\left(\vec{v'}\cdot\vec\nabla\Omega\right)r\sin\theta\vec{a_\varphi}= \\
        -\dfrac{1}{\rho_0}\vec\nabla p'-\vec\nabla\Phi'+\dfrac{\rho'}{\rho_0^2}\vec\nabla p_0{-\dfrac{\rho'}{4\pi\rho_0^2}\left(\vec\nabla\times\vec{B_0}\right)\times\vec{B_0}}\\ 
        {+\dfrac{1}{4\pi\rho_0}\left(\vec\nabla\times\vec{B'}\right)\times\vec{B_0}+\dfrac{1}{4\pi\rho_0}\left(\vec\nabla\times\vec{B_0}\right)\times\vec{B'}}
    \end{aligned}
,\end{equation}
\begin{equation}
    \left(\partial_t+\Omega\partial_\varphi\right)\left(\dfrac{\rho'}{\rho_0}-\dfrac{p'}{\Gamma_1 p_0}\right)+\vec{v'}\cdot\left(\vec{\nabla} \ln \rho_0-\dfrac{\vec{\nabla} \ln p_0}{\Gamma_1}\right)=0
,\end{equation}
\begin{equation}
    \begin{aligned}
        \partial_t \vec{B'}=\vec{\nabla}\times(\vec{v'}\times\vec{B_0})+\vec{\nabla}\times(\vec{v_0}\times\vec{B'})
    \end{aligned}
,\end{equation}
\begin{equation}
    \Delta \Phi' =4\pi G\rho'
,\end{equation}

with $\vec{v}$ the velocity field, $\rho$ the density, $p$ the pressure, $\vec{B}$ the magnetic field, $\Phi$ the gravitational potential, $\Gamma_1$ the adiabatic index, $G$ the gravitational constant, and $\vec{a}_i$ representing the unit vectors of the spheroidal coordinate basis, with the Einstein summation convention implied. The subscript $0$ denotes equilibrium quantities, while primes refer to Eulerian perturbations. 

Since the equilibrium structure is axisymmetric and stationary, the time and azimuthal dependencies of the perturbations are expressed as $e^{i(\omega t+\rm{ m}\varphi)}$, where $\omega$ is the mode angular frequency and $m$ is the azimuthal order. In the following, when developing Eqs. (3-7), we use the divergence-free property to substitute the first-order radial derivative of the radial component of the perturbated magnetic field as a function of the remaining terms in the divergence of $\vec{B'}$.

As in \citetalias{ouazzani_pulsations_2012}, we introduced the auxiliary variable $\mathrm{d}\Phi'$ and its associated equation, $\mathrm{d}\Phi'=\partial_\zeta\Phi'$, thereby rewriting Poisson’s equation from a second-order differential equation to two first-order ones. We also adopted the same change of variable, defining $\pi'=p'/\rho_0$.

We used the same dimensionless variables denoted by a tilde (see \citetalias{ouazzani_pulsations_2012}). We further nondimensionalized the magnetic field using

\begin{equation}
\begin{aligned}
    \tilde{\vec{B}}&=\sqrt{\dfrac{4\pi R^3}{M}}\dfrac{1}{R\Omega_{\mathcal{K}}}\vec{B}, 
\end{aligned}
\end{equation}

\noindent where $R$ is the star's radius, $M$ its mass, and $\Omega_{\mathcal{K}}$ the Keplerian angular frequency.
In what follows, we express all quantities in dimensionless form and omit the tilde for simplicity. 

\subsection{Purely toroidal magnetic fields}
\label{subsec:models_mag}

For the reasons presented in the introduction, and as a first step toward implementing more realistic magnetic fields, we considered a confined axisymmetric toroidal configuration. We chose a magnetic field in the following form:
\begin{equation}
    \vec{B_0}=B_0 B_{0\varphi}(r) \sin\theta \vec{e_\varphi},
\end{equation}
where $B_0$ is a normalization constant and $B_{0\varphi}$ the radial profile of the magnetic field. It is straightforward to show that this type of field is divergence-free as required. In this study, we used two types of magnetic fields, corresponding to two different analytical expressions for $B_{0\varphi}$.

\indent The first one is a Gaussian profile for $B_{0\varphi}$, with $r_0$ and $\sigma_0$ the parameters of the Gaussian (see Appendix \ref{subsec:app_mag1} for more details). This choice was made for testing purposes because this magnetic field is smooth, and by choosing $r_0$ and $\sigma_0$ appropriately we can avoid issues due to boundary conditions (by having a magnetic field going to zero at the surface and in the vicinity of the center). 

The second profile follows \cite{duez_relaxed_2010}, which provides a semi-analytic description of an axisymmetric mixed field with no rotation (for more details see Appendix \ref{subsec:app_mag2}). It is employed here for comparison purposes, particularly with the results of \citetalias{dhouib_detecting_2022}, regarding the magnetic extension of the TAR. In what follows, we consider only the toroidal component and restrict it to the radiative zone. Particular attention has been paid to the interface between the convective core and the radiative zone. The discontinuity is treated using double grid points with derivatives computed independently on the convective and radiative sides, while matching conditions ensure the continuity of the relevant physical quantities.

\section{Numerical method}
\label{sec:numerical_method}

We used the method presented in \citetalias{ouazzani_pulsations_2012}, which consists of a multi-domain spectral expansion onto spherical harmonics (hereafter SH) series for the angular components of the eigenfunctions and a radial differentiation scheme \citep{scuflaire_liege_2008} well adapted to sharp variations in the structure of evolved stars for the radial part of the eigenfunctions. The system of ODEs and boundary conditions constitute a two-point boundary eigenvalue problem (BVEP) solved using a Newton-like method. In this extension, we implemented the case of a toroidal magnetic field, keeping the code as modular as possible to ease future extensions to other magnetic field configurations and additional physical processes. Here, we detail the modifications made to the numerical method to account for the new system of equations and its specificities, compared to the original \acor formulation. We then present the modular framework developed to automate the construction of the eigenvalue problem and its resolution. A flowchart describing the whole procedure is available in Fig. \ref{fig:MACO_mindmap}. Finally, we present tests assessing the convergence and stability of the code.

\subsection{Spectral expansion}

We performed a similar spectral expansion to that in \citetalias{ouazzani_pulsations_2012}. The main difference in our approach lies in the treatment of the vector equations, whereby we used the vector spherical harmonics (hereafter VSH) \citep{rieutord_linear_1987} instead of a poloidal-toroidal decomposition as done in \citetalias{ouazzani_pulsations_2012}, to limit the complexity of the system of equations. The VSH are defined as follows: 

\begin{equation}
    \begin{aligned}
        \vec{\mathcal{R}_\ell^m}&=iY_\ell^m\vec{a_\zeta},\\
        \vec{\mathcal{S}_\ell^m}&=i\left(\partial_\theta Y_\ell^m \vec{a_\theta}+\dfrac{1}{\sin\theta}\partial_\varphi Y_\ell^m \vec{a_\varphi}\right),\\
        \vec{\mathcal{T}_\ell^m}&=\left(\dfrac{1}{\sin\theta}\partial_\varphi Y_\ell^m \vec{a_\theta}-\partial_\theta Y_\ell^m \vec{a_\varphi}\right)\\
    \end{aligned}
.\end{equation}
Any perturbed vector field, $\vec{V'}$, is then expressed as
\begin{equation}
    \label{eq:BVSH}
    \vec{V'}=\sum_{ \ell\geq |m|}^{+\infty}\left(u_{V \ell,m}(\zeta)\ \vec{\mathcal{R}_\ell^m}+v_{V \ell,m}(\zeta)\ \vec{\mathcal{S}_\ell^m}+w_{V \ell,m}(\zeta)\ \vec{\mathcal{T}_\ell^m}\right)
,\end{equation}
where $u_{V \ell,m},\ v_{V \ell,m},$ and  $w_{V \ell,m}$ are the corresponding radial and transverse components of the vector field, $\vec{V'}$.

\subsection{Simplifications and symmetries}
\label{subsec:simplifications_symetries}

By considering an axisymmetric toroidal magnetic field that is either symmetric or antisymmetric with respect to the equator, and using the symmetries of the equilibrium structure, we can separate the system of equations for each parity with respect to the equator and for each azimuthal order. Therefore, for a given value of $m$, we can split the eigenvalue problem into two independent sets of ordinary differential equations (ODEs) coupling the spectral coefficients, one with $\ell$ of the same parity as $m$ (even mode), and the other with $\ell$ of the opposite parity as $m$ (odd mode).

Therefore for a symmetric magnetic field, for a mode of azimuthal order $m$, decomposed on $M$ SH, we have $\forall j \in [1,M]$,
\begin{equation}
\begin{aligned}
    &\ell=|m|+2(j-1)+par \ \rm{for} \ \pi_\ell',\rho_\ell',\Phi_\ell',d\Phi_\ell',u_{v,\ell},v_{v,\ell},u_{b,\ell},v_{b,\ell},\\ 
    &\ell_p=|m|+2(j-1)+1-par \ \rm{for} \ w_{v,\ell_p}, w_{b,\ell_p}.
\end{aligned}
\label{eq:landlp}
\end{equation}
Here we recall that the quantities $X_\ell'$ are the spectral coefficients corresponding to the quantity $X'$. Even modes (same parity as $m$) have $par = 0$, whereas odd modes (otherwise) have $par=1$. The variables $\ell$ and $\ell_p$ are different angular degrees of the SH. This corresponds to the same description used in \citetalias{ouazzani_pulsations_2012} with the addition of the spectral coefficients related to $B'$. 

The case of magnetic fields without specific symmetry with respect to the equator is also possible (at the expense of simplicity), but is not presented here.

\subsection{Boundary conditions}
\label{subsec:BC}
The choice of a purely axisymmetric toroidal field permits to reduce the system of equations to a 4M first-order ODEs system. Therefore to solve the eigenvalue problem, only four boundary conditions are needed. For the boundaries at the surface and exterior to the star, because we consider confined magnetic fields, no changes are required. By contrast, the magnetic field has an impact on the boundary conditions at the center. We extend the regularity conditions described in \citepalias{ouazzani_pulsations_2012} to the case of toroidal magnetic fields. However, this approach failed to ensure the regularity of the magnetic field perturbation, despite having tested various sets of equations. To overcome this issue, which is related to the chosen set of equations and not to the underlying assumptions, we decided to consider only magnetic fields with a strong decay close to the center so that the innermost layers remain unaffected by the magnetic field. We shall tackle this issue in a future study.

\subsection{Matrix of the problem}
For the numerical method used to solve the eigenvalue problem, we used the same formalism as in \citetalias{ouazzani_pulsations_2012} (described in more detail in \citealt{ouazzani_phd_17-03_rmouazzanipdf_2011}), with some minor changes to make it more resilient to changes in the system of equations. We defined $\sigma$ as the angular frequency of oscillation normalized by the Keplerian angular frequency. We employed a Newton-like iterative method: starting from an initial guess, $\sigma_0$, we computed the closest eigenfrequency, $\sigma$. This yields a correction, $\delta\sigma$, to the initial guess, $\sigma_0$, and we took $\sigma=\sigma_0+\delta\sigma$ as a new guess. The procedure was repeated until convergence was achieved.

Following \citetalias{ouazzani_pulsations_2012}, the system of equations is therefore described as follows:

\begin{equation}
    \left\{
        \begin{aligned}
            (D_{11} +\delta\sigma D_{12}&)d_\zeta \vec{y_1}+(D_{21}+\delta\sigma D_{22}) d_\zeta \vec{y_2} = \\ & (A_{11}+\delta\sigma A_{12})\vec{y_1}+(A_{21}+\delta\sigma A_{22})\vec{y_2},\\
            0 = & (B_{11}+\delta\sigma B_{12})\vec{y_1}+(B_{21}+\delta\sigma B_{22})\vec{y_2},
        \end{aligned}
    \right.
\end{equation}
where $\vec{y}_1$ and $\vec{y}_2$ are column vectors containing the spectral coefficients of the unknowns
\begin{equation}
\label{eq:eq_variables}
    \vec{y}_1=\left[
\begin{matrix}
   \pi_{\ell}'\\ 
   \Phi_{\ell}' \\
   d\Phi_\ell'\\
   u_{v\ell}'
\end{matrix}\right]
\ \ \ \ \ \ \ \ \ \mathrm{and}  \ \ \ \  \vec{y}_2=\left[\begin{matrix}
   v_{v\ell}'\\ 
   w_{v\ell_p}'\\ 
   \rho_\ell'\\
   u_{B\ell}'\\
   v_{B\ell}'\\ 
   w_{B\ell_p}'
\end{matrix}\right]
,\end{equation}
with $\ell$ and $\ell_p$ defined in Eq. (\ref{eq:landlp}). The matrices $D_{11}$, $D_{12}$, $D_{21}$, and $D_{22}$ are the ones whose coefficients are multiplied by the first-order radial derivative of the unknowns. Note that these matrices differ from those derived in \citetalias{ouazzani_pulsations_2012}. The matrices $A_{11}$, $A_{12}$, $A_{21}$, $A_{22}$, $D_{11}$, $D_{12}$, $D_{21}$, and $D_{22}$ correspond to the following differential equations: 

\begin{center}
\begin{equation}
\label{eq:eq_derivative}
    \left[\begin{matrix}
       \hbox{Momentum equation along}\,  \vec{\mathcal{R}_\ell^m} \\
        \hbox{Definition of}\,  \rm{d}\Phi_\ell' \\
        \hbox{Poisson's Equation} \\
        \hbox{Continuity Equation} 
    \end{matrix}\right]
,\end{equation}
\end{center}

whereas  $B_{11}$, $B_{12}$, $B_{21}$, and $B_{22}$ correspond to the following algebraic equations: 

\begin{equation}
\centering
    \left[\begin{matrix}
        \hbox{Momentum equation along } \vec{\mathcal{S}_\ell^m} \\
        \hbox{Momentum equation along } \vec{\mathcal{T}_{\ell_p}^m} \\
        \hbox{Adiabatic relation } \\
        \hbox{Induction equation along } \vec{\mathcal{R}_\ell^m} \\
        \hbox{Induction equation along } \vec{\mathcal{S}_\ell^m} \\
        \hbox{Induction equation along } \vec{\mathcal{T}_{\ell_p}^m} \\
    \end{matrix}\right]
\label{eq:algebric_eq}
.\end{equation}

Using the system of equations (\ref{eq:algebric_eq}), we can replace the components of $\vec{y}_2$ with their expressions in terms of $\vec{y}_1$, and rearrange the system of equations (\ref{eq:eq_derivative}), so that only $\vec{y}_1$ unknowns are differentiated. Then, we can invert $(D_{11}+\delta\sigma D_{12})$ (if $\delta\sigma$ is sufficiently small; see \citetalias{ouazzani_pulsations_2012}), recovering a system similar to \citetalias{ouazzani_pulsations_2012} and proceed accordingly.

\subsection{Modular code with SymPy}
\label{subsec:modular_SymPy}

Rather than extending the \acor code in its original implementation, we chose to restructure it into a modular framework in which the full set of equations can be directly specified by the user while the underlying  numerical method remains unchanged. To make the analytical treatment of these equations easier, we implemented a symbolic pipeline based on the Python library SymPy \citep{sympy}, allowing us to derive the full system of MHD oscillation equations in a two-dimensional stellar context. The inherent complexity of these equations, particularly after their expansion on the spheroidal coordinate system, and projection onto the SH or VSH basis, makes manual derivation both cumbersome and prone to error. The automation enabled by this pipeline overcomes these issues by simplifying and minimizing errors in the derivation process.

\begin{figure}[h]
    \includegraphics[width=0.9\columnwidth,trim=0cm 0.8cm 0cm 0.8cm, clip]{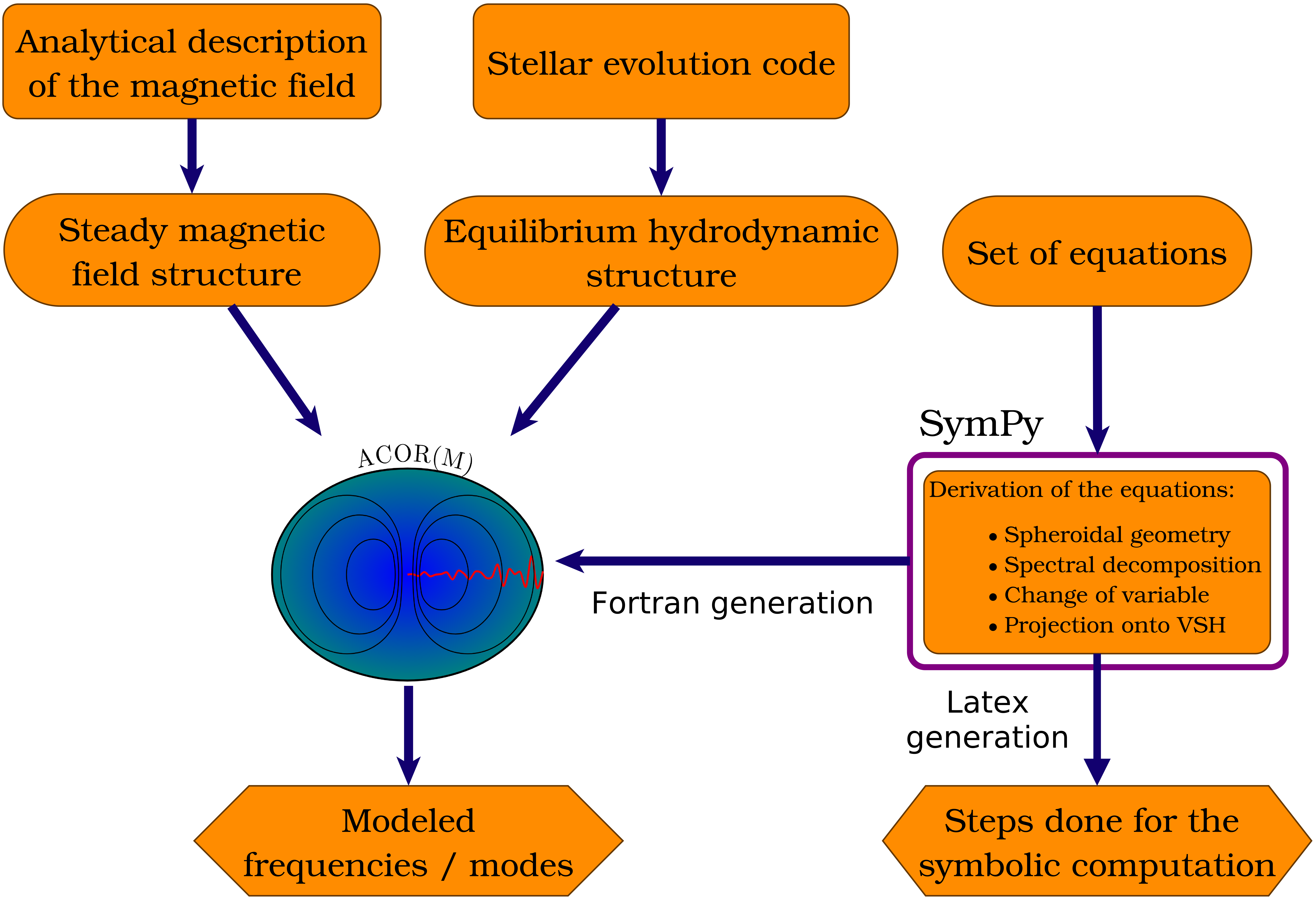}
    \caption{Overview of \acor.  A 1D or 2D stellar model from a stellar evolution code, a steady magnetic field configuration from an analytical description, and a system of equations are taken as inputs, then the analytical development is processed by the SymPy pipeline, and leads to the eigenvalue problem being solved and the mode frequencies being computed.}
    \label{fig:MACO_mindmap}
\end{figure}

In this pipeline, all physical quantities and differential operators are defined symbolically. The MHD equations are then automatically expanded and projected using SymPy's symbolic manipulation capabilities, including the application of relevant simplifications and variable transformations. The resulting output is a coupled system of ODEs for the spectral coefficients, which is subsequently prepared for numerical discretization via the automatic generation of Fortran code, to be used as input to the \acor solver. In addition, the pipeline produces a LaTeX file that documents the full derivation of the system of equations, providing a transparent and traceable record of the analytical development. A summarized pipeline flowchart is shown on the right-hand side of Fig. \ref{fig:MACO_mindmap}.

This symbolic approach offers several key advantages. It significantly reduces the risk of algebraic mistakes, particularly in the treatment of Lorentz force terms and geometric couplings, and it accelerates model development by allowing physical assumptions, such as the magnetic field configuration or rotation profile, to be changed without manual derivation. It also improves consistency and reproducibility, as all expressions are generated from a controlled set of definitions and rules. This pipeline has already proven effective: in addition to correcting minor errors in previous formulations, it successfully reproduces the oscillation spectra obtained with the original \acor code, thereby confirming both its accuracy and robustness. In the current version, only the case of a toroidal magnetic field has been implemented, but this modular approach will facilitate the future inclusion of more complex magnetic field configurations or additional physical processes.

\subsection{Convergence and performance} 

In this subsection, we assess the convergence of the code based on the case of $\gamma$ Dor stars. The stellar models used in this article were computed using the Cesam2K20 stellar evolution code \citep{morel_cesam_1997,morel_cesam_2008,Marques2013,Manchon2025} for a $1.4$ solar-mass star aged $20$Myr at the zero-age main sequence (ZAMS), whose main parameters are given in Table \ref{table:model_key} (model \textit{ce4k} unless otherwise specified, see Appendix \ref{subsec:stellar_models} for more details).

\begin{table}
\caption{Main characteristics of the models \textit{ceXk}.}
\label{table:model_key}
\centering
\begin{tabular}{c c}
\hline\hline
Model name & \textit{ceXk} \\
\hline\hline
$M/\mathrm{M}_\odot$& $1.40$ \\
$T_{\mathrm{eff}}$ ($\mathrm{K}$) & 6761 \\
$\log$ $L /\mathrm{L}_\odot$ & 0.642 \\
$\log$ $g$($\mathrm{cm}\cdot\mathrm{s}^{-2}$) & 4.25 \\
Age (Myr) & $20$ \\
$R/\mathrm{R}_\odot$ &  1.52\\
$R_\mathrm{c}/\mathrm{R}_\odot$ & $6.13\cdot 10^{-2}$\\
$\rho_\mathrm{c}$~($\mathrm{g}\cdot\mathrm{cm}^{-3}$) & $75.0$ \\
Number of grid points & $X\cdot10^3$ \\
\hline\hline
\end{tabular}

\tablefoot{Mass, effective temperature, luminosity, surface gravity, age, radius, convective core radius, central density, and number of points in the model (from top to bottom).}

\end{table}

\begin{figure}[h]
    \centering
    \includegraphics[width=\columnwidth,trim=0cm 0.1cm 0cm 0.1cm, clip]{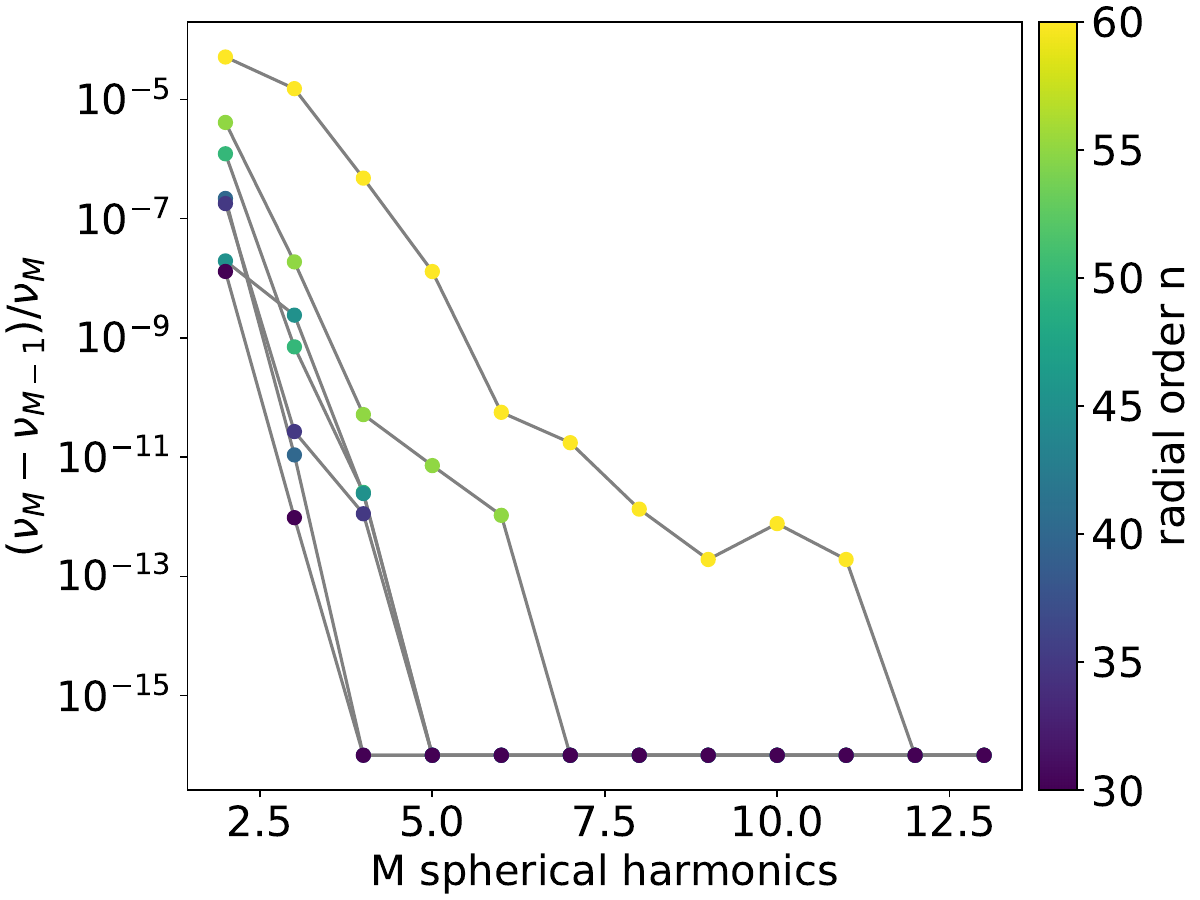}
    \caption{Relative variation in the frequency, $\nu$, obtained with $M-1$ and $M$ SH, as a function of the number, $M$, of SH, for prograde dipolar modes with radial orders from $n=30$ to $60$ at $B_0=3000$~kG for the model \textit{ce4k}. A lower boundary is fixed at the machine precision $10^{-16}$ to remain visible in the log scale when it equals 0.X}
    \label{fig:convergence_SH}
\end{figure}

Rotation and magnetic field break the spherical symmetry of the stellar structure, and cause the eigenfunctions to deviate from a single spherical harmonic. Multiple SHs become coupled in the mode spectral decomposition and accordingly the number of SH required to precisely compute the modes increases as rotation becomes faster and/or magnetic field strength stronger. As underlined in \citetalias{ouazzani_pulsations_2012}, there is no automatic and totally reliable method to identify modes, particularly in terms of angular degree. In this study, we track individual modes by slowly increasing the magnetic field strength. Our study was based mainly on prograde dipolar g modes ($m=1$, $l=1$) with radial orders from $n=30$ to $n=60$ representative of $\gamma$ Dor asteroseismic observations \citep{li_period_2019,li_gravity-mode_2020} with a magnetic field peaking at $3000$~kG, which corresponds to the upper bound of the strong-field regimes inferred in red-giant cores \citep{deheuvels_strong_2023,villate_seismic_2026}. For $\gamma$ Dor stars, we expect these values to be lower due to flux conservation. However, these observational constraints concern poloidal configurations. We therefore deliberately retained strong toroidal magnetic field strengths to test the code in the high-field regime. 
To assess the precision with respect to the number of SH, we computed a set of modes with different truncation levels for the SH series ($M$ being the number of SH used for the spectral expansion) for a given value of magnetic field strength $B_0$. This is illustrated in Fig. \ref{fig:convergence_SH}, where the relative difference between calculations with two consecutive numbers of SH $(\nu_M-\nu_{M-1})/\nu_M$ is below $10^{-10}$ for all the modes studied once $M\geq6$. This indicates that six SHs are sufficient to compute precisely enough the oscillations for these modes at this magnetic field strength.

\begin{figure}[h]
    \centering
    \includegraphics[width=\columnwidth,trim=0cm 0.1cm 0cm 0.1cm, clip]{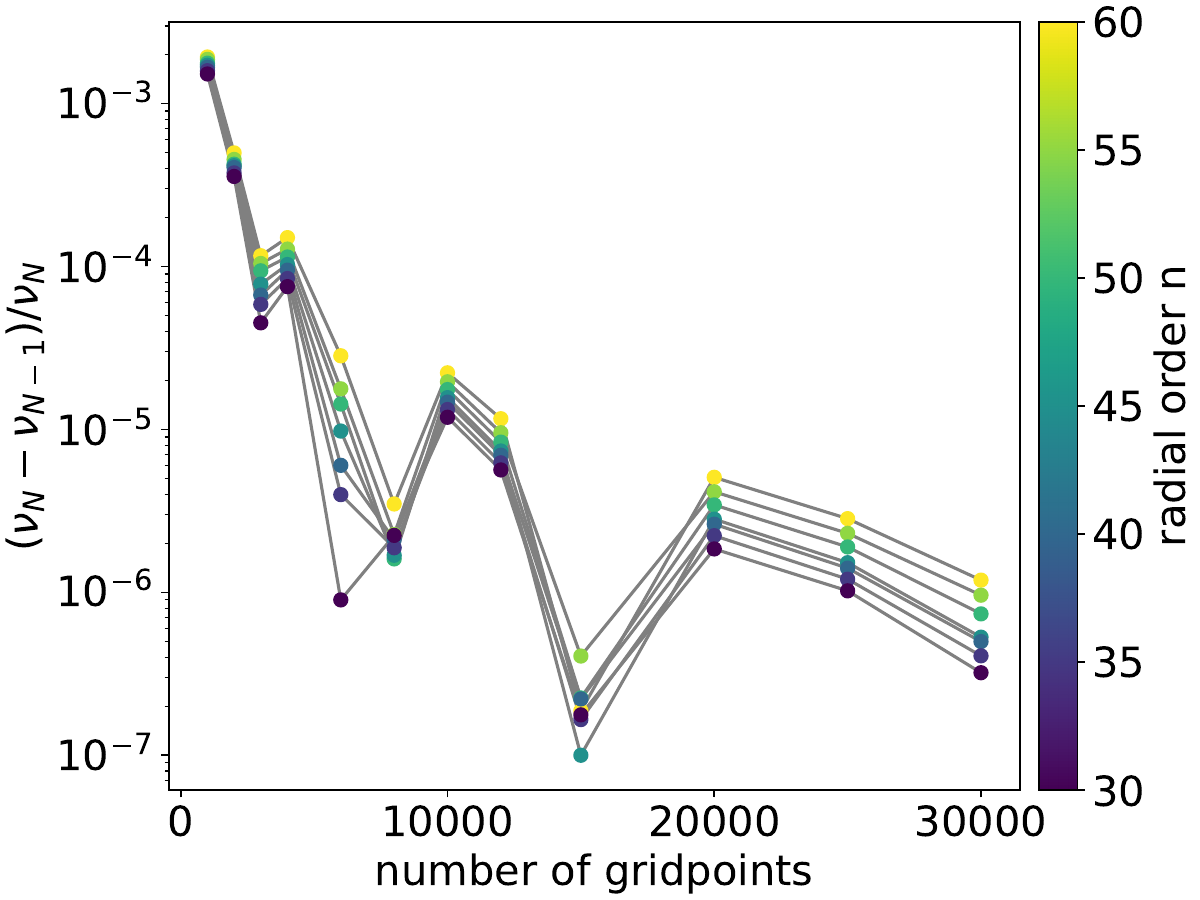}
    \caption{Relative variation in the frequency, $\nu$,  performed with consecutive radial resolutions, N and N-1, for prograde dipolar modes with radial orders from $n=30$ to $60$ at $B_0=3000$~kG. The models used are the \textit{ceXk}, where \textit{X} is a number varying from \textit{0.5} to \textit{30} (see Section \ref{subsec:stellar_models}).}
    \label{fig:convergence_radial}
\end{figure}

Radial resolution also plays an important role in the convergence of the computed frequencies. Fig. \ref{fig:convergence_radial} shows the relative difference between frequencies computed at successive radial resolutions $|\nu_{N}-\nu_{N-1}|/\nu_N$. The difference decreases as the model resolution increases. For a resolution of 4000 grid points, the difference is below $\simeq 10^{-5}-10^{-6}\mu$Hz, depending on the mode studied. This resolution is adopted in the remainder of this work, as it provides a good compromise between the precision of the computed frequencies and the computational cost. The  sharp variations visible in Fig. \ref{fig:convergence_radial} result from a grid-point distribution optimized for the stellar structure rather than the g modes studied here. This radial sampling directly impacts the computed frequencies, leading to varying discrepancies between successive radial resolutions.

It can be noted from both Fig. \ref{fig:convergence_SH} and Fig. \ref{fig:convergence_radial} that convergence depends on the mode studied. Higher-radial-order modes are more affected by the magnetic field, and therefore require a better resolution, and a greater number of SHs to achieve the same level of precision as their low-radial-order counterparts.
 
The computational time scales similarly to \citetalias{ouazzani_pulsations_2012}: it depends on the model resolution, the number, $M$, of SHs used in the spectral decomposition, and the system of equations used (different systems can be used depending on the physics chosen, see Section \ref{subsec:modular_SymPy}). An example of the numerical resources used by this new implementation compared to the initial one is available in Table~\ref{table:comparison_comptime} for M$=6$~SH and a rotation rate of $\Omega/2\pi=20\mu$Hz for a scan in frequency from $25$ to $30\mu$Hz with a resolution of $0.25\mu$Hz. The higher computational time and memory usage arise from the larger system of equations, as well as the more expanded expressions produced by the modular SymPy-generated formulation.

\begin{table}
\caption{Numerical resources used by \acor.}
\label{table:comparison_comptime}
\centering
\hspace{-0.5cm}
\setlength{\tabcolsep}{2pt}
\begin{tabular}{c c c c c}
\hline\hline
Version & Magnetic (kG) & Matrix size & Time & Peak memory\\
 &  & & & usage\\
\hline\hline
\citetalias{ouazzani_pulsations_2012} & No & $5.9\cdot 10^6$ & 1m14s & 0.48GB \\
This study & $0$~kG & $9.8\cdot 10^6$ & 2m59s & 0.88GB \\
This study & $500$~kG & $9.8\cdot 10^6$ & 3m56s & 0.88GB \\
\hline\hline
\end{tabular}
\tablefoot{The matrix size of the problem to solve corresponds to a system of four first-order ODEs and a system of three algebraic equations for \citetalias{ouazzani_pulsations_2012} or six algebraic equations for this version. These are band matrices, so their size reported here corresponds to the number of elements in their band. Benchmarks were performed on a laptop equipped with an Intel Core i7-13700H processor (up to 5.0 GHz) and 32 GB RAM. The code was compiled with gfortran 13.3.0 with optimization level -O3.}
\end{table}

\section{Comparison with the perturbative approach}
\label{sec:perturbative}

Comparing the results of \acor with those obtained using perturbative approaches serves a double purpose. Firstly, the aim is to demonstrate that in the slow rotation and low magnetic field strength regime, the complete calculations converge on the perturbative ones. In the absence of a code similar to \acor in the community that would allow us to treat both rotation and magnetism in a non-perturbative way, this is one way to validate the current implementation. Secondly, such a comparison helps to define the domain of validity of the existing perturbative methods.

In this section, we first present the methods used to obtain the perturbative frequency shifts (with and without rotation), and then compare them with the frequencies computed with \acor, applied to 1D stellar models (using the model \textit{ce4k} unless otherwise specified) with the toroidal magnetic field resulting from the description used in \citetalias{dhouib_detecting_2022}. Although the stellar models were computed in 1D assuming uniform rotation, \acor is capable of handling 2D models with differential rotation. This choice is made for simplicity and to facilitate a comparison with perturbative and TARM approaches.

\subsection{Nonrotating case}
\label{subsec:pert_no_rot}
When both rotation and magnetic fields are absent, each eigenfunction of the unperturbed problem, denoted 
$\xi_0$, can be written in terms of a single SH \citep{unno_nonradial_1989}:

\begin{equation}
\boldsymbol{\xi}_0 = 
\left[
\xi_r Y_\ell^m(\theta, \varphi),\,
\xi_h \partial_\theta Y_\ell^m,\,
\xi_h \frac{im}{\sin\theta} Y_\ell^m
\right]e^{i(\omega_0 t+m\varphi)}
,\end{equation}
where $\xi_r$ and $\xi_h$ are the radial and horizontal components of the eigenfunction, $\vec{\xi}_0$, respectively, and $\omega_0$ is the angular frequency of the mode.

We defined $<.,.>$ as the usual inner product: 
\begin{equation}
\langle f, g \rangle = \iiint \vec{f^*} \cdot \vec{g} \, \rho \, r^2 \sin\theta \, dr \, d\theta \, d\varphi\
.\end{equation}

For a sufficiently weak magnetic field, the frequency change remains small compared to the spacing between unperturbed modes and does not significantly alter the eigenfunctions. Thus, a perturbative treatment is possible to compute $\delta\nu_{\rm pert}$ the frequency shift and is given by \citep{gough_effect_1990}

\begin{equation}
\label{eq:deviation_mag}
\delta \nu_{\rm{pert}} = 
\frac{1}{4\pi \omega_0} 
\frac{
\langle \vec{\xi}_0, \vec{\delta F}_{\mathrm{L}}  \rangle
}{
\langle \vec{\xi}_0 , \vec{\xi}_0 \rangle
},
\end{equation}
with
\begin{equation}
\begin{aligned}
\label{eq:lorentz_forces_pert}
\vec{\delta F}_{\mathrm{L}} =& \frac{-1}{4\pi\rho_0} \Big[ (\vec\nabla \times \vec{B}_0) \times  \vec{\delta B} + (\vec\nabla \times  \vec{\delta B}) \times \vec{B}_0 \\
+&\dfrac{\vec{\nabla}\cdot(\rho_0 \vec{\xi}_0)}{\rho_0}\left(\vec\nabla\times\vec{B}_0\right)\times\vec{B}_0 \Big] ,
\end{aligned}
\end{equation}
where the first two terms correspond to the linearized Lorentz force, while the last term accounts for the compressibility associated with the mode.
The Lagrangian perturbation of the magnetic field was derived from the linearized induction equation: 
\begin{equation}
\vec{\delta B} = \nabla \times (\boldsymbol{\xi}_0 \times \mathbf{B}_0)
.\end{equation}

We followed modes with radial orders ranging from $n=30$ to $n=60$, in steps of 5, while progressively increasing $B_0$ for the model \textit{ce4k}. The frequency shift, $\delta\nu$, corresponds to that computed by \acor. For weak magnetic fields, examining the relative frequency shift difference between the two methods highlights how well the complete approach matches the perturbative approach.

\begin{figure}[h]
    \centering
    \includegraphics[width=\columnwidth,trim=0cm 0.1cm 0cm 0.1cm, clip]{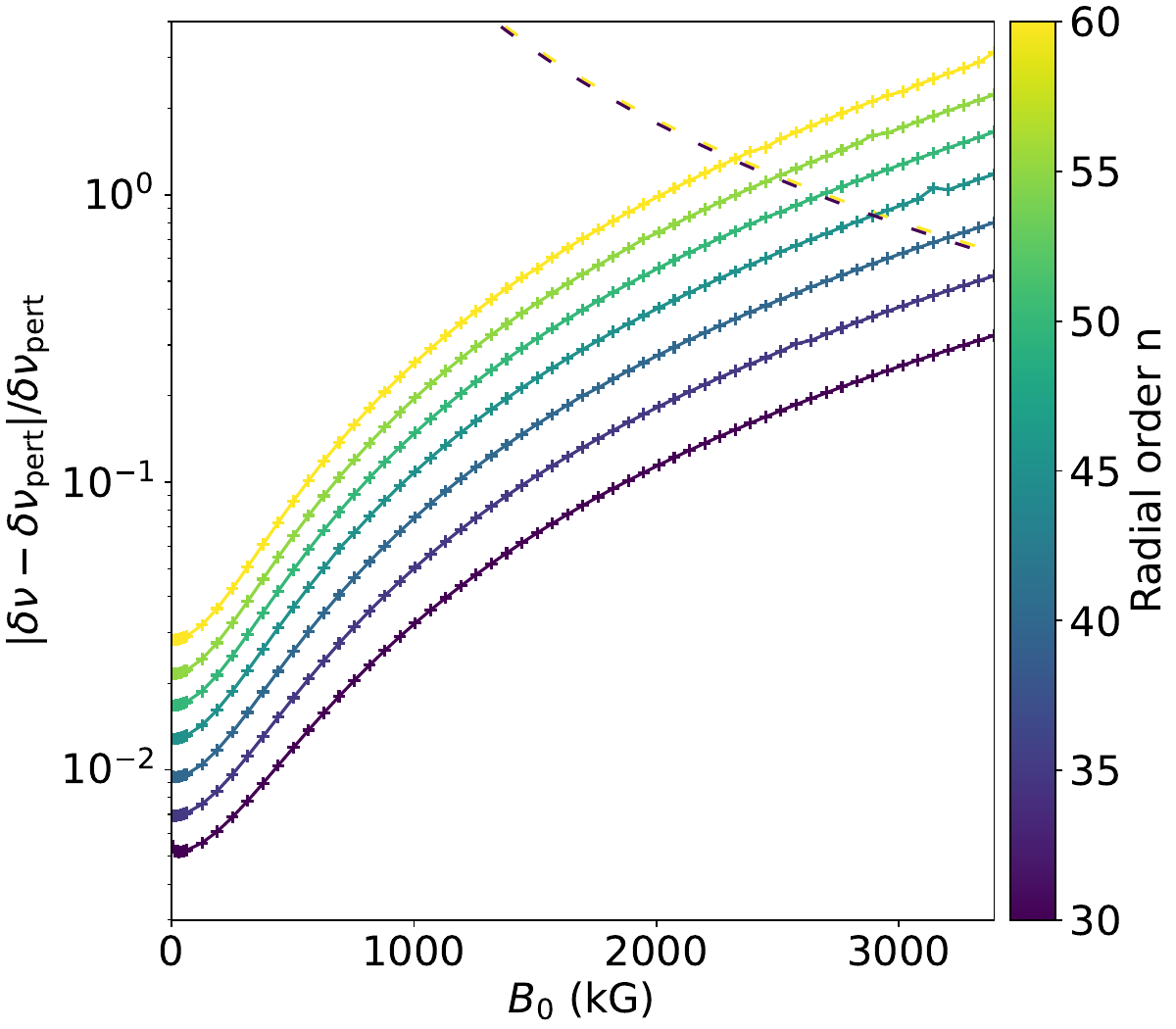}
    \caption{Evolution of the relative frequency-shift difference between the perturbative approach, $\delta\nu_{\rm{pert}}$, and  \acor $\delta\nu$ along the magnetic field strength, $B_0$. The colors represent the same modes as in \ref{fig:pert_norot_1}. A value of 0 means that it perfectly follows the perturbative approach. The dashed lines are the relative frequency-shift difference when it is assumed that the difference frequency shift equals the theoretical frequency resolution, $\nu_{\it Kepler}$.}
    \label{fig:pert_norot_2}
\end{figure}

\begin{figure}[h]
    \centering
    \includegraphics[width=0.9\columnwidth,trim=0cm 0.1cm 0cm 0.1cm, clip]{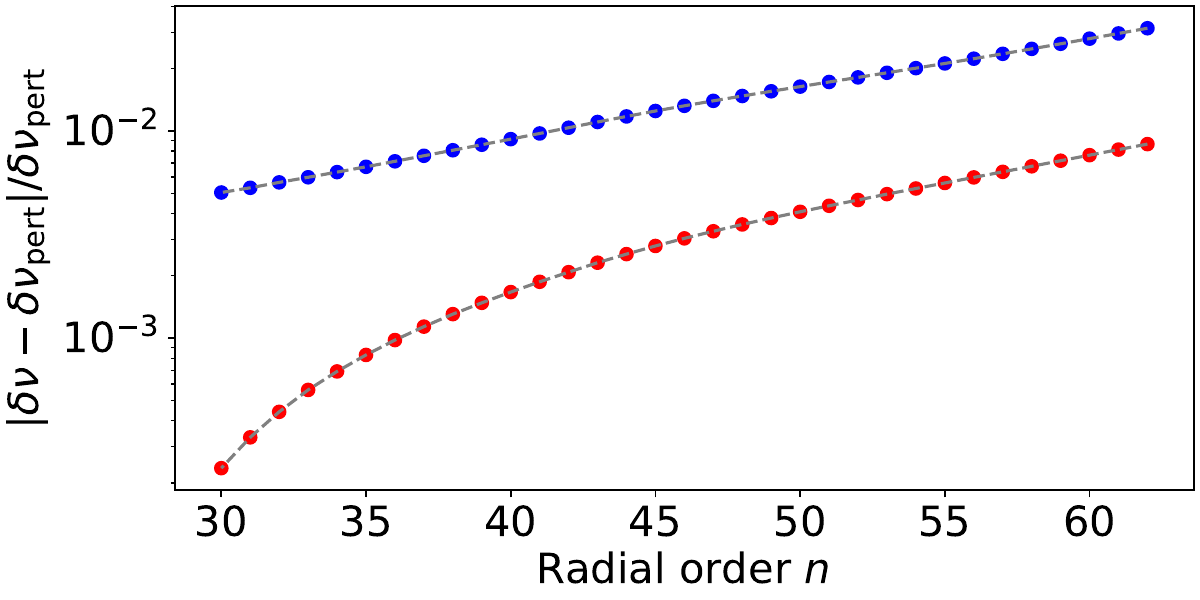}
    \caption{Relative frequency-shift difference along the radial order of the modes. Both panels are for $B_0=5$~kG. The blue points are modes computed with the model \textit{ce4k} with 4000 grid points in the radial direction, while the red points were computed with the model \textit{ce8k} with 8000 grid points.}
    \label{fig:pert_norot_3}
\end{figure}

Fig. \ref{fig:pert_norot_2} shows that the relative frequency-shift difference converges on zero and showcases a good agreement between the two methods at low field strengths. However, the ratios do not exactly reach 0 but rather remain slightly above, with a small residual discrepancy. We investigated the source of discrepancy by looking at different resolutions. For the same value of $B_0$, Fig. \ref{fig:pert_norot_3} shows that the radial resolution has an impact on the frequency computation, where the relative frequency-shift difference between the two methods is closer to zero for the high-resolution model. This also explains why high-radial-order modes deviate more from the perturbative results at low $B_0$:  due to their stronger radial variations, a better resolution is required to achieve the same level of precision as their low radial counterparts.

For stronger magnetic fields, in Fig. \ref{fig:pert_norot_2}, the relative frequency-shift difference reaches a value of 1 around $B_0\simeq2020$~kG for the mode $n=60$. This implies that the perturbative estimate of the frequency shift, $\delta\nu_{\rm pert}$, is only half of the shift computed with \acor, highlighting the breakdown of the perturbative approximation in this regime. In the same figure, the dotted lines indicate the case in which the relative frequency-shift difference is evaluated by assuming that the frequency shift difference equals the theoretical frequency resolution \citep{kallinger_freq_2007} from the {\it Kepler} mission, $\nu_{\rm{{\it Kepler}}}=1/(4T_{\rm\it Kepler})\simeq2.0\times 10^{-3}\mu {\rm Hz}$ ($T_{\rm\it Kepler}\simeq$ four years of observations). Although in practice, frequencies may be determined more precisely than $\nu_{\rm{{\it Kepler}}}$ when no unresolved modes contaminate the spectral window, it remains a useful reference for estimating the frequency shift required to be detectable in observations. When the discrepancies between the methods exceed $\nu_{Kepler}$, they become observationally significant, setting a practical validity for the perturbative treatment. For instance, for the mode $n=60$, this limit is reached around $B_0\simeq 2330$~kG and is not even reached for the mode $n=30$ for $B_0 < 3000$~kG.

\begin{figure}[h]
    \centering
    \includegraphics[width=\columnwidth,trim=0cm 0.1cm 0cm 0.1cm, clip]{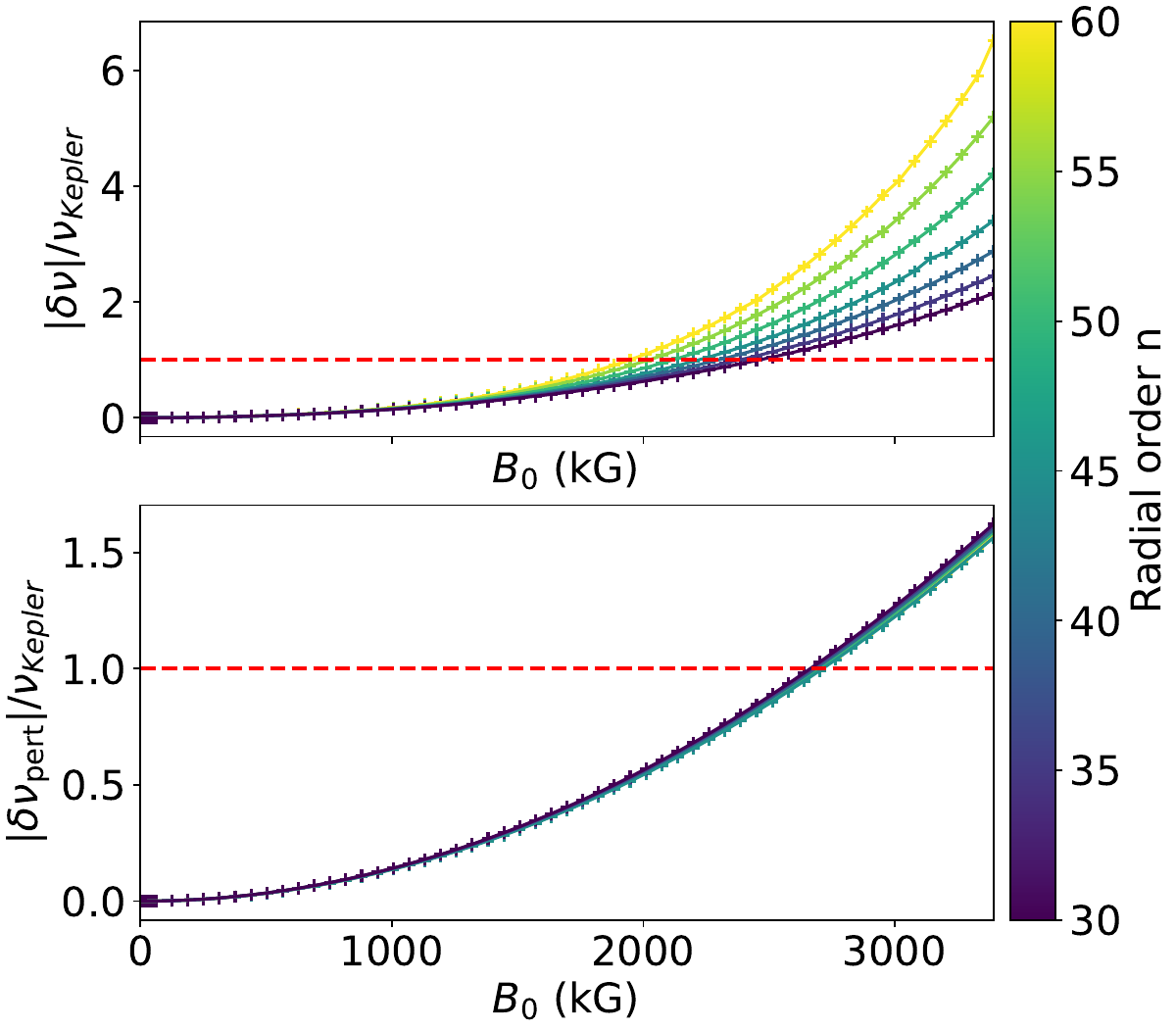}
    \caption{Evolution of the frequency shift with increasing magnetic field strength, $B_0$. The color corresponds to modes of different radial orders ($n=30$ in blue to $60$ in red, incremented by 5). The \textit{upper panel} corresponds to the magnetic frequency shift, $\delta\nu$, computed by \acor over the theoretical resolution frequency of the {\it Kepler} mission, $\nu_{\rm{{\it Kepler}}}$. The \textit{lower panel} corresponds to the magnetic frequency shift computed with the first-order perturbative approach, $\delta\nu_{\rm pert}$, over $\nu_{\rm{{\it Kepler}}}$. The dashed red line corresponds to a frequency shift equal to $\nu_{\rm{{\it Kepler}}}$.}
    \label{fig:pert_norot_1}
\end{figure}

Fig. \ref{fig:pert_norot_1} presents the evolution of the magnetic frequency shifts, $\delta\nu$ and $\delta\nu_{\rm{pert}}$, normalized by $\nu_{Kepler}$, as a function of the magnetic field strength, $B_0$. In the upper panel, the normalized frequency shift $\delta\nu/\nu_{\rm \textit{Kepler}}$ exceeds unity at $B_{0}\simeq2000$~kG for the mode with the highest radial order ($n=60$). In the \acor appproach, this characteristic field strength depends on the radial order, where the normalized frequency shift for lower-$n$ modes exceed unity for stronger magnetic fields. In contrast, within the perturbative framework, the corresponding field strength at which the normalized shift reaches unity occurs at larger $B_0$, and has a weaker radial order dependence.

To investigate this weak radial order dependence, we used the splitting coefficient (which differs slightly from \citet{hasan_probing_2005}) defined by
\begin{equation}
    S_c=\dfrac{\delta\nu}{B_0^2}
.\end{equation}

\begin{figure}[h]
    \centering
\includegraphics[width=\columnwidth,,trim=0cm 0.1cm 0cm 0.1cm, clip]{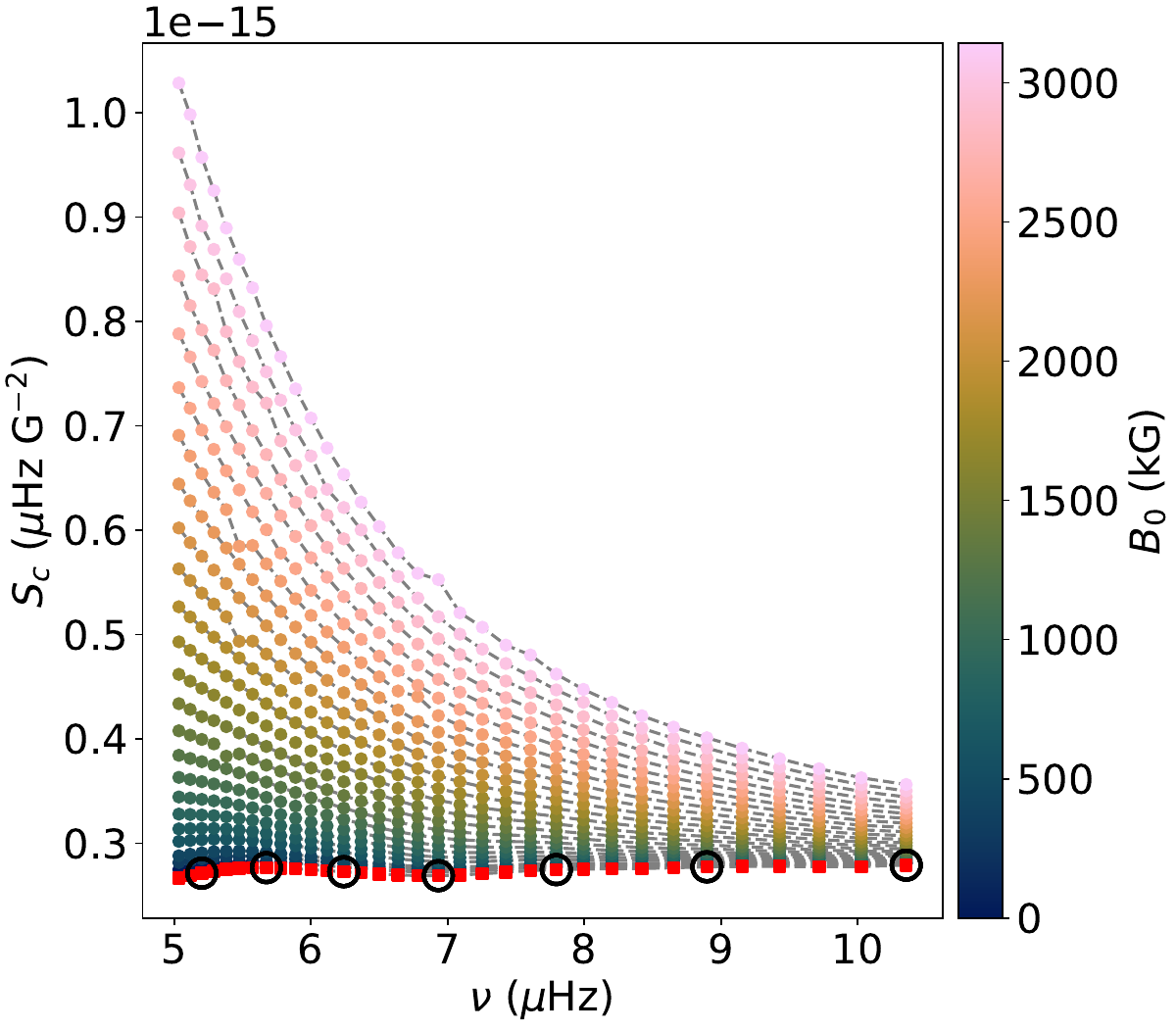}
    \caption{Splitting coefficient, $S_c$, along the radial order of the mode. The colored circles were computed using the \acor code, whereas the red squares were computed through the perturbative approach. The color of the circles represents the magnetic field strength used to compute the frequencies. The encircled red squares are the modes from $n=30$ to $n=60$ in steps of 5 in radial order.}
    \label{fig:pert_Sc}
\end{figure}

Splitting coefficients are by definition independent of $B_0$ for the perturbative approach, whereas this is not the case in the non-perturbative one. This permits to illustrate the progressive breakdown of the perturbative approach. 
The results are represented in Fig. \ref{fig:pert_Sc}, for the modes $n=30$ to $n=62$ with increasing magnetic field strengths. The red squares represent the splitting coefficients, $S_{\rm c}$, from the perturbative approach. We observe that the lower the magnetic field, the closer we are to the splitting coefficients, $S_c$, computed with the perturbative approach. Thus, at low magnetic field strengths the results from \acor are correctly converging on the ones predicted by the perturbative approach.

\begin{figure}[h]
    \centering
\includegraphics[width=\columnwidth,,trim=0cm 0.1cm 0cm 0.1cm, clip]{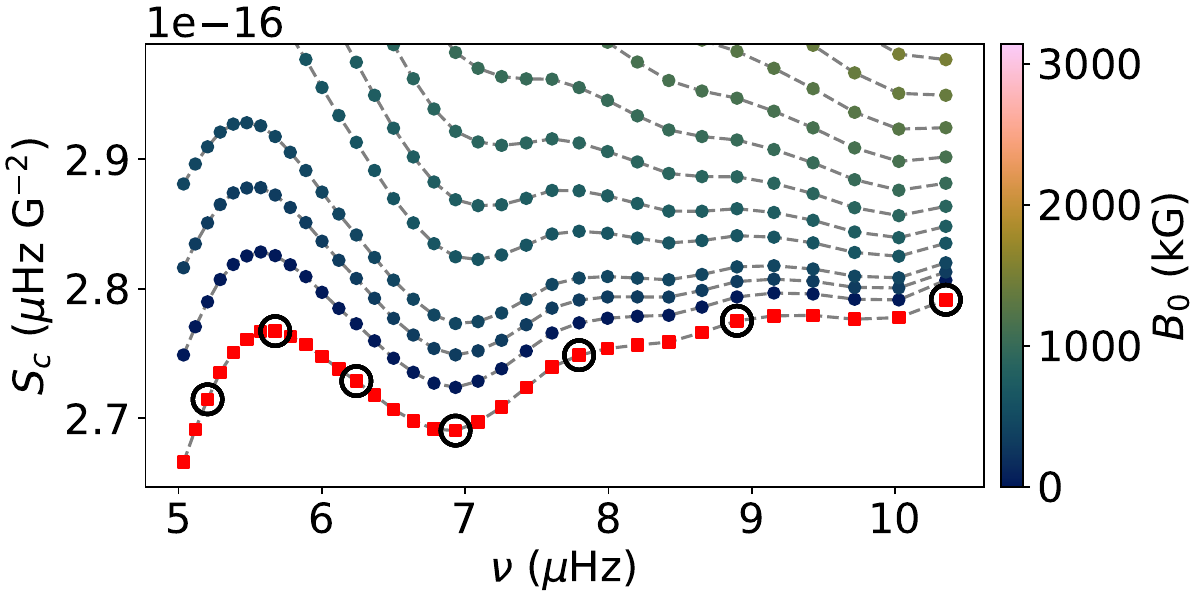}
    \caption{Zoom on Fig. \ref{fig:pert_Sc} around the $S_c$ computed using the perturbative approach represented by the red squares.}
    \label{fig:pert_Sc_zoom}
\end{figure}

The splitting coefficients computed with \acor increase with magnetic field strength, whereas they remain constant in the perturbative approach. For strong magnetic fields, the difference in splitting coefficients between modes of different radial order becomes significant: on the order of approximately $10^{-15}\mu$Hz $\mathrm{G}^{-2}$ between the mode $n=30$ and $n=60$ in the \acor computations, compared to approximately only $10^{-17}\mu$Hz $\mathrm{G}^{-2}$ in the perturbative case (see Fig.~\ref{fig:pert_Sc_zoom}). This explains the much weaker radial-order dependence on the frequency shift obtained with the perturbative approach compared to the \acor results.

The breaks seen in Fig. \ref{fig:pert_Sc} (for instance, for the mode $n=45$ around $\nu_0=5.5\mu$Hz for a $B_0=1600$~kG and $2200$~kG) are due to avoided crossings with modes of higher $\ell >1$. When passing the avoided crossing, we have to jump to the other branch to continue tracking the same mode along $B_0$ creating this break in the curve. 

Another point of concern has been identified in Fig. \ref{fig:pert_norot_2}, where the frequency splittings obtained through the perturbative approach do not follow the expected $1/\nu_0$ dependence predicted for toroidal fields. This is discussed in Appendix \ref{sec:termbyterm}. 

For magnetic field strengths higher than $B_{0}\simeq3500$~kG, the mode of radial order $n=60$ becomes difficult to follow due to the growing number of avoided crossings with modes of higher angular degree. Therefore, its frequency can no longer be computed reliably. The inherent limits of the frequency computations will be discussed in detail in Section \ref{subsec:limits}.

\subsection{Case with slow rotation}

\begin{figure*}[!htp]
    \centering
    \subfigure{\includegraphics[width=0.90\columnwidth,trim=0cm 0.1cm 0cm 0.1cm, clip]{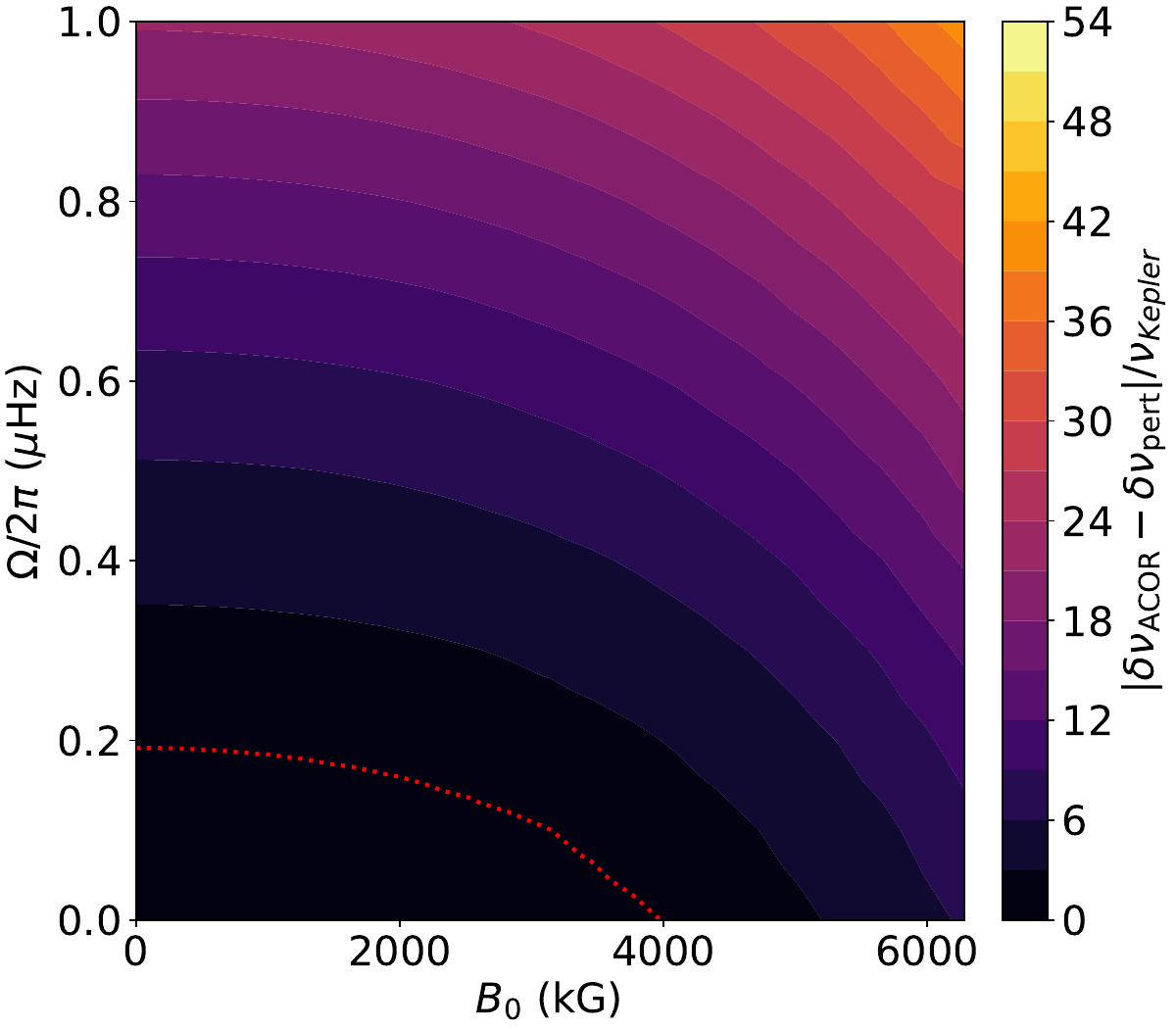}}
    \subfigure{\includegraphics[width=0.90\columnwidth,trim=0cm 0.1cm 0cm 0.1cm, clip]{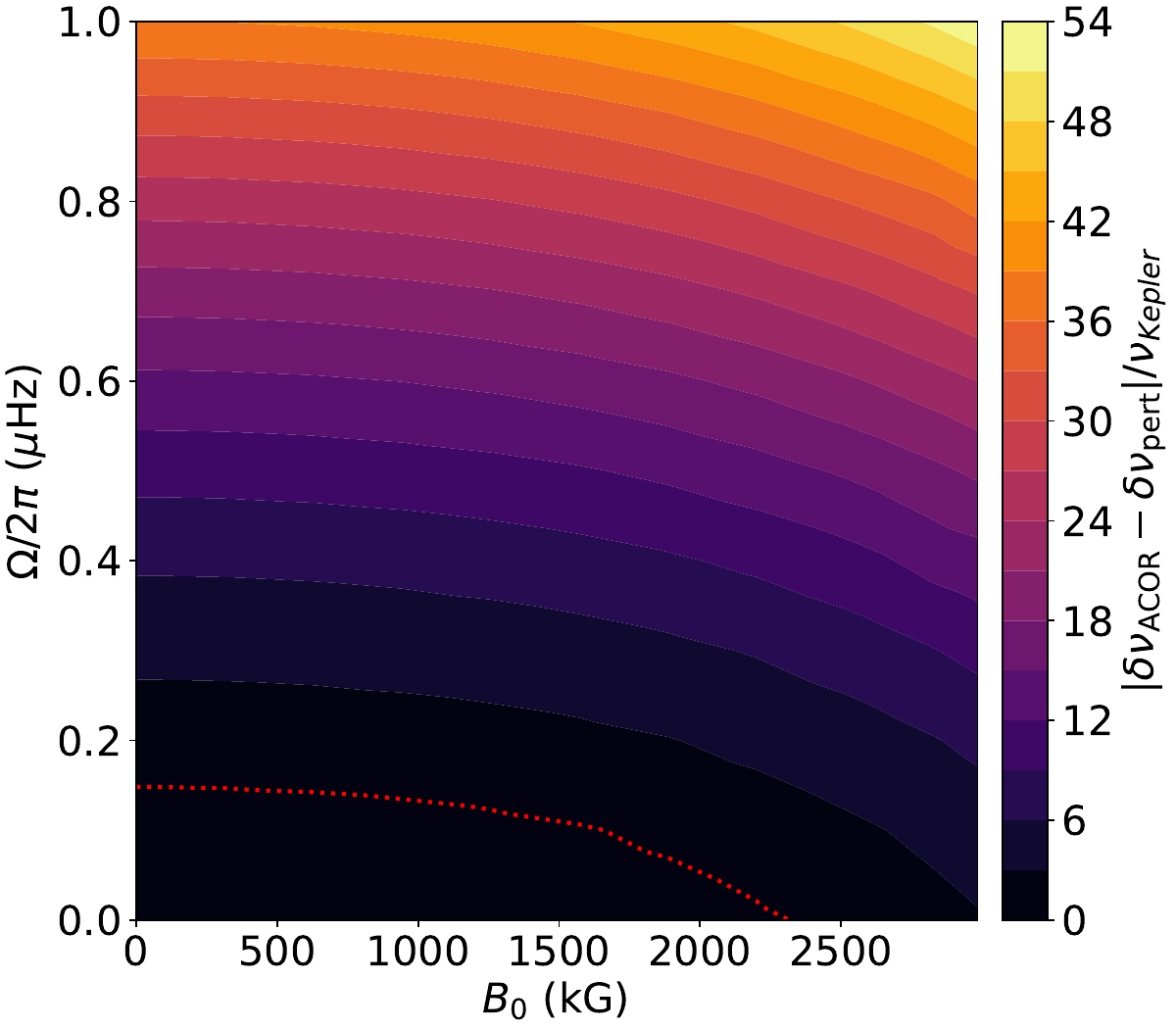}}
    \caption{Difference in frequency shifts between $\delta\nu$ and $\delta\nu_{\rm pert}$ over the theoretical frequency resolution of the {\it Kepler} mission $\nu_{\rm{{\it Kepler}}}=0.002\mu$Hz for the mode $n=30$ (\textit{left}) and $n=60$ (\textit{right}). The x axis corresponds to $B_0$ in kG and the y axis the rotation frequency in microhertz. The dotted red line corresponds to the case where the difference in frequency shifts is equal to the {\it Kepler} theoretical frequency resolution. Note that the $B_0$ range differs between the two panels.}
    \label{fig:mode_pert2D}
\end{figure*}

Since \acor has been designed to handle both rotation and toroidal magnetic field, we were interested in their cumulative effects. We compared these results with the first-order perturbative approach of rotation and magnetic field. 
Including rotation through a first-order perturbative approach only slightly modifies Eq.~(\ref{eq:deviation_mag}) by adding a second term related to rotation \citep{unno_nonradial_1989}: 

\begin{equation}
\delta \nu_{\mathrm{pert}} = 
\frac{1}{4\pi \omega_0} 
\frac{
\langle \vec{\xi}_0, \vec{\delta F}_{\mathrm{L}}  \rangle
}{
\langle \vec{\xi}_0 , \vec{\xi}_0 \rangle
}+\dfrac{\langle (m\Omega-i\vec{\Omega}\times)\vec{\xi}_0 , \vec{\xi}_0 \rangle}{\langle \vec{\xi}_0 , \vec{\xi}_0 \rangle}
.\end{equation}

We explore a regime of very low rotation, with a rotation frequency below $1\mu$Hz, corresponding to a rotation period longer than  11.6 days. Figure~\ref{fig:mode_pert2D} shows the deviation between the magnetic frequency splitting computed with \acor and the first-order perturbative expression, normalized to the {\it Kepler} frequency resolution, $\nu_{\rm{{\it Kepler}}}$. For low magnetic field strengths and low rotation rates, the shifts computed with \acor and with the perturbative approach show good agreement. As expected, the deviation increases with both the magnetic field strength, $B_0$, and angular velocity, $\Omega$. The dotted red line indicates the limit, in terms of $\Omega/2\pi$ and $B_0$, above which the discrepancy exceeds the {\it Kepler} resolution. The comparison also reveals that the sensitivity to both magnetic and rotational effects is greater for the high-order mode ($n=60$), as its deviation grows faster than for the lower-order mode ($n=30$). High rotation rates lead to larger deviations as the magnetic field increases (and conversely), showing that the effects of rotation and magnetism are correlated rather than independent, since both jointly modify the restoring forces and the spatial structure of the modes. \acor reproduces the results obtained with the perturbative approach at low magnetic field strengths and shows the breakdown of the perturbative approach for a magnetic field higher than $\simeq2000$~kG, depending on the mode studied in the case of our model.

\section{Comparison with the \citet{dhouib_detecting_2022} TARM approach}
\label{sec:dhouib}

\begin{figure*}[!htp]
    \centering
    \subfigure{\includegraphics[width=0.98\columnwidth,trim=0cm 0.1cm 0cm 0.1cm, clip]{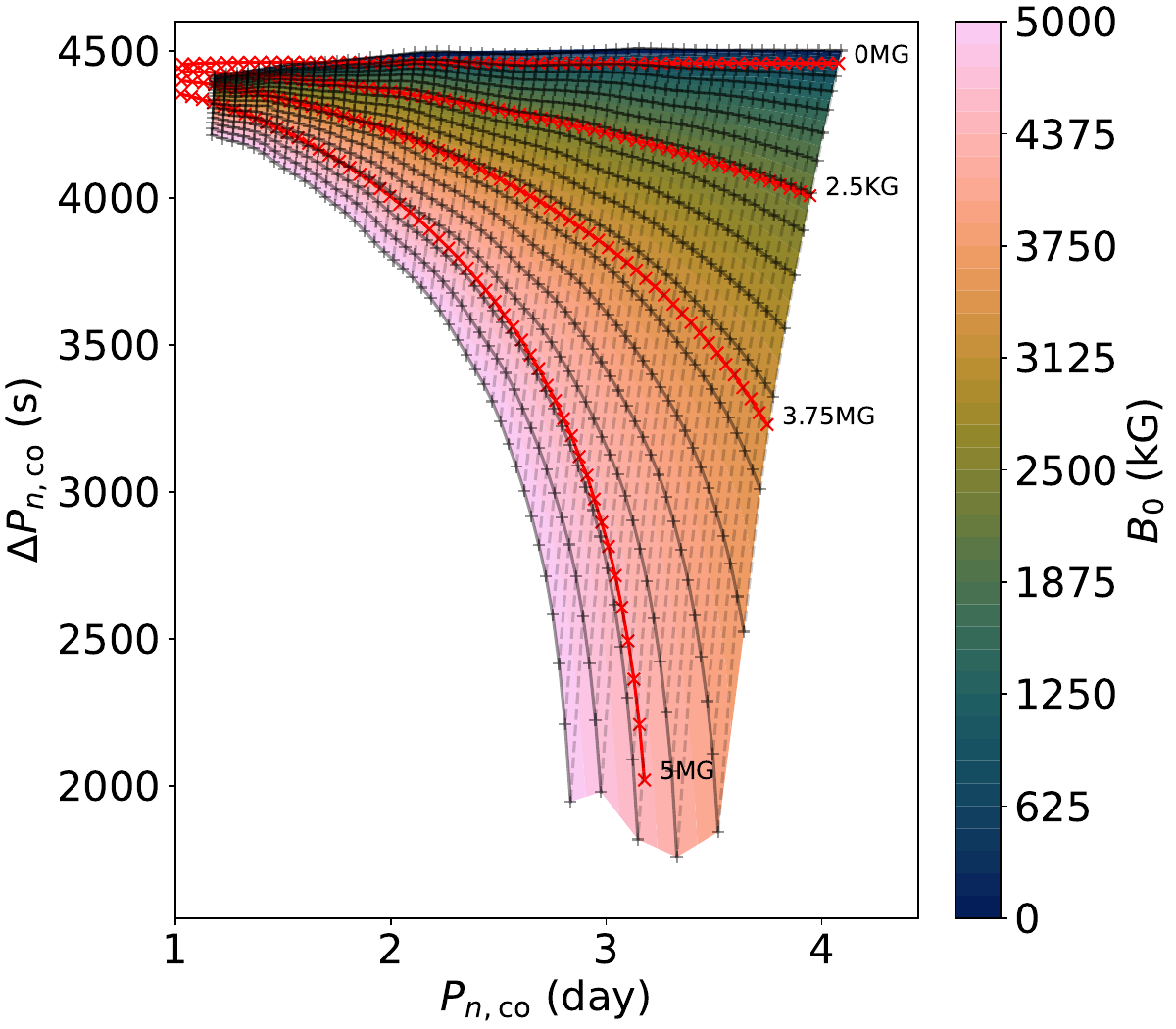}}
    \subfigure{\includegraphics[width=0.98\columnwidth,trim=0cm 0.1cm 0cm 0.1cm, clip]{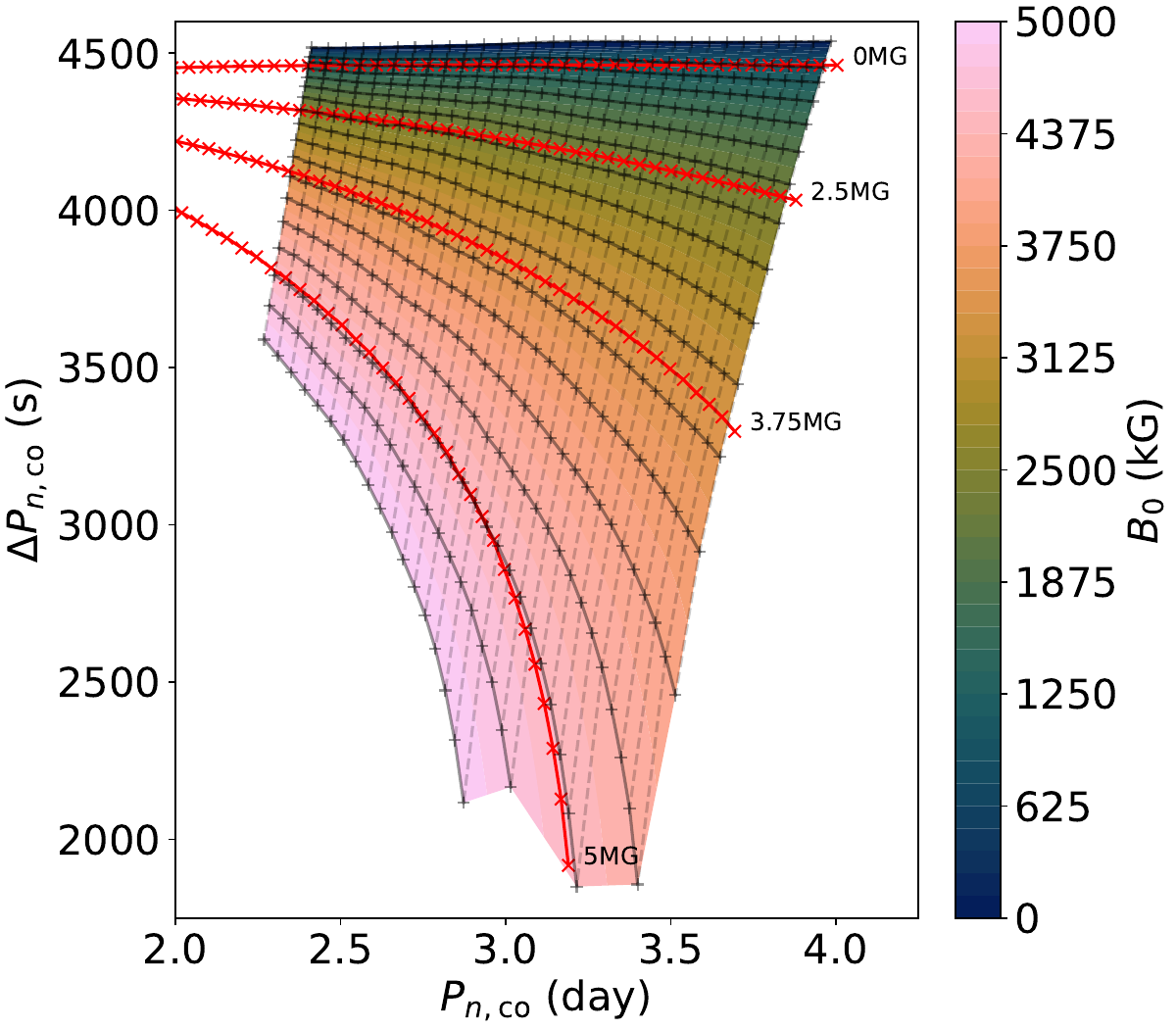}}
    \caption{Period-spacing patterns as a function of period in the corotating frame for dipolar prograde inertial modes with the \textit{ce4k} model with $\Omega/2\pi=20\mu$Hz (\textit{left}) and $\Omega/2\pi=10\mu$Hz (\textit{right}). The black lines are the period-spacing patterns computed with \acor for different values of $B_0$, ranging from $0$~to~$5000$~kG in steps of $250$~kG and the color map corresponds to their interpolated values. The dashed black lines correspond to the period spacing with the same radial order, n, but different values of $B_0$. The red curves correspond to the period-spacing patterns computed with the TARM from \citetalias{dhouib_detecting_2022} for different values of $B_0$, indicated at the end of each curve in kilogauss.}
    \label{fig:dhouib_iso}
\end{figure*}

To test the validity of our new implementation of \acor for strong toroidal magnetic fields and high rotation rates, we compare our results to the work of \citetalias{dhouib_detecting_2022}. That work introduces a new extension of the TAR that accounts for the effects of a magnetic field, leading to the so-called TARM. In addition to our assumptions, the study \citetalias{dhouib_detecting_2022} adopts the Cowling approximation (neglecting the perturbation of the gravitational potential) and omits the terms associated with the compressibility of the mode. It also relies on the Jeffreys-Wentzel-Kramers-Brillouin (JWKB) approximation, valid when the eigenfunctions vary much more rapidly in the radial direction than the background structure (see \citet{unno_nonradial_1989}). Finally, the asymptotic frequencies are derived by applying the radial quantization to the dispersion relation.
To make the comparison as accurate as possible, we used the same code\footnote{During the course of the comparison with \citet{dhouib_detecting_2022}, an error was identified in their magnetic field intensities, which are $2\pi\sqrt{4\pi}$ stronger than what is reported. Once this mistake is corrected, both methods give similar results.} as in \citetalias{dhouib_detecting_2022} and generated, with the TARM, the pulsation spectra of the same stellar model as for \acor, for identical rotation rates and magnetic field strengths. The same magnetic topology as \citetalias{dhouib_detecting_2022} was assumed, described in Appendix \ref{subsec:app_mag2}. The comparisons were performed with the model \textit{ce4k} in the nonrotating case or with a rotation of $\Omega/2\pi=10\mu$Hz and $\Omega/2\pi=20\mu$Hz (respectively, $\simeq15\%$ and $\simeq30\%$ of the Keplerian angular frequency, or in terms of the rotation period $\simeq 1.16$~days and $\simeq 0.58$~days). The frequencies were computed for radial orders starting at $n=80$ and extending down to a few radial orders below the frequency ridge of the $m=-1$, $\ell=3$ modes (whenever permitted by the numerical and physical constraints discussed in Section~\ref{subsec:limits}). This lower radial order corresponds to $n=49$ for $\Omega/2\pi=10\mu$Hz and $n=24$ for $\Omega/2\pi=20\mu$Hz.
To investigate the effect of the magnetic field, we used the period-spacing  diagrams, $\Delta P_{n}$ versus period $P_{n}$. The period spacing is defined as the period difference between modes of consecutive radial orders, $n$, with the same angular degree, $l$, and azimuthal order, $m$. For pure g modes in the asymptotic regime, it is expected to be nearly constant \citep{tassoul_1980} and rotation introduces a slope in this pattern in the inertial frame of reference \citep{bouabid_effects_2013}. In the corotating frame, the period spacing, $\Delta P_{n,\mathrm{co}}$, as a function of $P_{n,\mathrm{co}}$ is roughly constant within the TAR, so that magnetism is one of the processes responsible for these deviations. Figure \ref{fig:dhouib_iso} shows the results of the comparison in the case of $\Omega/2\pi=20\mu$Hz and $10\mu$Hz. Although the tendencies are similar in terms of period spacing, an offset is visible even in the nonmagnetic case. As a sanity check to investigate this offset, we performed comparisons between \acor and TOP \citep[Two-dimensional Oscillation Program][]{reese_modelisation_2006,reese_pulsation_2009} in the case without magnetic field, and we found almost no differences (with an error in $\Delta P_n$ below $\simeq0.1$ s). Therefore, for the nonmagnetic case, these discrepancies are due to the different approximations made in \citetalias{dhouib_detecting_2022} (Cowling, JWKB, incompressibility, asymptotic expression).

A grid of period-spacing patterns computed with \acor for different magnetic field strengths was built. In order to estimate the error made when determining magnetic field strengths using the TARM, a best-fit algorithm was then used to determine the equivalent magnetic field strength from \acor calculations that best reproduces the period spacing obtained from \citetalias{dhouib_detecting_2022}.

The results are summarized in Table \ref{table:best_fit_B0}. We selected only high radial order modes to better comply with the assumptions made in \citetalias{dhouib_detecting_2022} (asymptotic expression valid for high radial orders). The differences between \acor and the TARM approach increase slightly with the rotation rate and with the intensity of the magnetic field. The differences can also be traced back to the offset in the nonmagnetic case. For $B_0=2500$~kG, the difference in $B_0$ between the perturbative approach and \acor is around $651$~kG for the highest radial order mode, whereas here (see Fig. \ref{fig:erroB0pert}) it is only around $113$~kG between the TARM approach and \acor.

Overall, given the qualitative agreement, this validates the correct implementation of the code and provides indicative bounds on the errors introduced by the TARM approach, on the order of $10\%$. Larger errors are expected in less stratified regions, where the assumptions of the TAR begin to break down. 

\begin{table}
\caption{Estimated magnetic field strength, $B_0$, obtained by fitting the period-spacing pattern, $\Delta P_n$, computed with the TARM from \citetalias{dhouib_detecting_2022}.}
\label{table:best_fit_B0}
\centering
\begin{tabular}{c cc}
\hline\hline
Input $B_0$ (kG) & \multicolumn{2}{c}{Best-fit $B_0$ (kG) in the \acor grid} \\
\hline
& $\Omega/2\pi = 10\,\mu$Hz & $\Omega/2\pi = 20\,\mu$Hz \\
\hline
500  & $2387 \pm 17$   & $2319 \pm 17$ \\
750  & $3426 \pm 4$   & $3393 \pm 5$ \\
1000 & $4560 \pm 10$ & $4496 \pm 24$ \\
\hline\hline
\end{tabular}
\tablefoot{The fitting was done using the period-spacing pattern grid computed with \acor (model \textit{ce4k}). For each fit, the ten highest radial order modes were used. Uncertainties correspond to the standard deviation derived from these ten modes.}
\end{table}

\section{Discussions and limitations}
\label{sec:discussions}

We first discuss the difficulties encountered when computing pulsations in certain frequency ranges. We then place our results in the context of the existing magnetic field measurements, in red giant and $\gamma$ Dor stars.

\subsection{Limits of the frequency computation}
\label{subsec:limits}

\begin{figure}[h]
    \centering
    \includegraphics[width=\columnwidth,trim=0cm 0.1cm 0cm 0.1cm, clip]{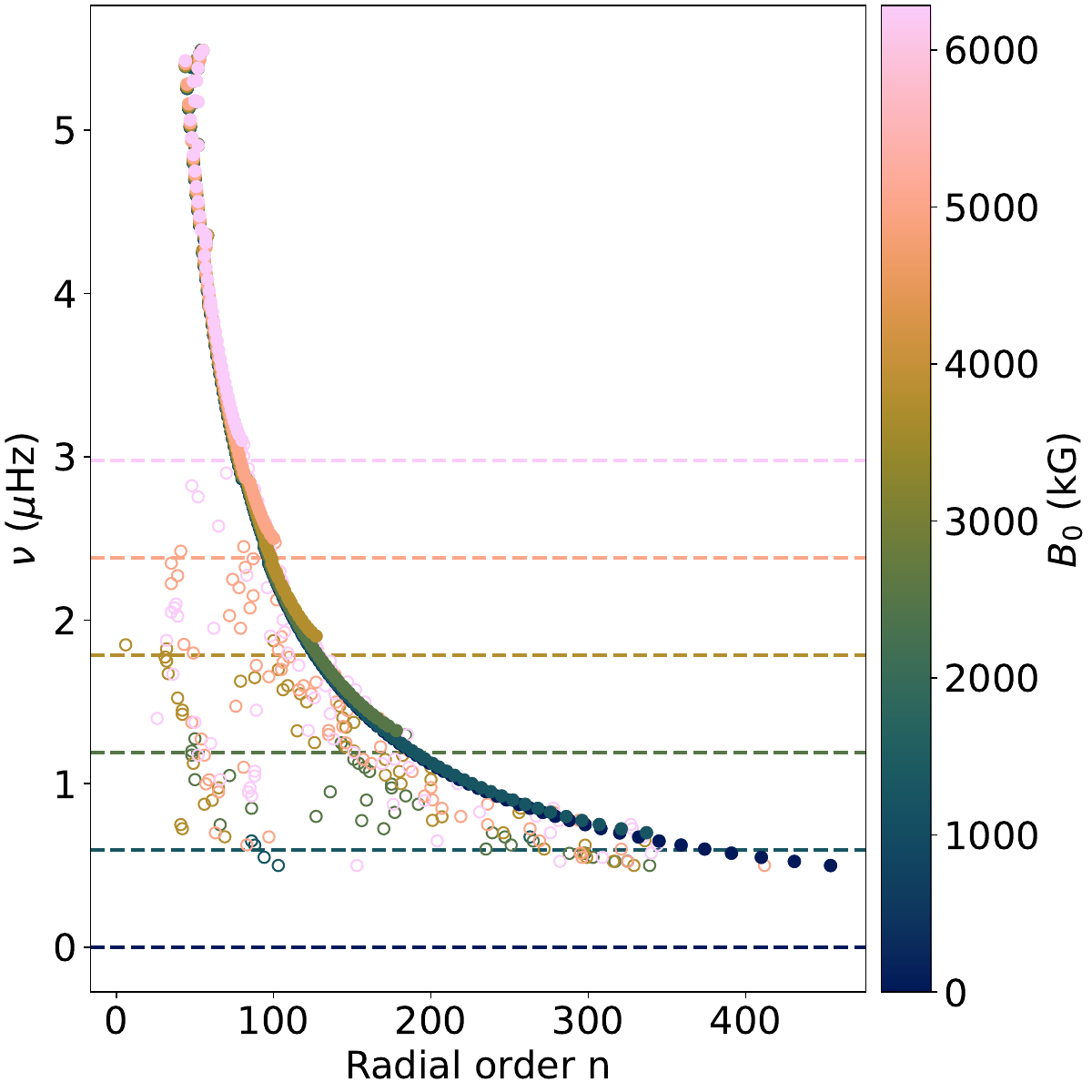}
    \caption{Frequency, $\nu$, of the dipolar prograde modes in the corotating frame with a rotation, $\Omega/2\pi=10\mu$Hz, from a scan between $10\mu$Hz and $15\mu$Hz with a resolution of $0.02\mu$Hz as a function of the radial order, $n$, for different magnetic field strength. Non-converging modes or those exhibiting spurious discontinuities are shown as empty circles. The dashed lines correspond to the Alfvén frequency for each magnetic field strength. We use the model \textit{ce4k} with a Gaussian magnetic field centered around $r_0=0.15$ and $\sigma_0=0.01$.}
    \label{fig:conv_freq}
\end{figure}

Although \acor can compute modes over a broad frequency range, it encounters convergence issues or irregular solutions for the eigenfunctions in certain regions due to numerical effects. This problem can also appear in the nonmagnetic case. With rotation, modes tend to organize into ridges in the $\Delta P$-$P$ plane with the same angular degree, $\ell$. Along such ridges, the radial order increases rapidly as the frequency decreases, eventually exceeding the resolution capabilities of the model. Such a ridge is visible in the nonmagnetic case in Fig. \ref{fig:conv_freq} for prograde dipolar modes.

Within the same frequency range, lower-$\ell$ modes can still exist, but they become difficult to identify for two main reasons. Firstly, when a low-$\ell$ mode lies in a dense frequency region of high-$\ell$ modes, it becomes hard to find. Secondly, mode coupling through avoided crossings can further complicate identification: low-$\ell$ modes couple to higher-$\ell$ modes, and this makes them hard to track. If the resolution is not fine enough, coupling with high-radial-order high-$\ell$ modes can even lead to numerical issues. 

A similar situation arises in nonrotating or slowly rotating models with strong magnetic fields. In the magnetic configurations considered here, high-$\ell$ mode frequencies increase more rapidly with field strength than low-$\ell$ modes. As a result, low-degree modes become embedded in a dense spectrum of high-degree ones, leading to the same identification and resolution issues described above.

For rotating stars, low-$\ell$ modes close to the rotation rate are found isolated from the high-$\ell$ modes and the problems described before should not happen. But as the magnetic field strength increases, modes close to or below the Alfvén frequency, $\nu_A$, often fail to converge, or display spurious features in their eigenfunctions. We define the Alfvén frequency as
\begin{equation}
    \nu_A=\dfrac{B_0}{2\pi\sqrt{4 \pi \rho_0}r}
.\end{equation}
In Fig. \ref{fig:conv_freq}, we scanned a frequency range of a few microhertz above the rotation frequency for different magnetic field strengths, $B_0$. Plotting the mode frequencies against radial order reveals ridges down to a critical value of $\nu$, below which modes either do not converge or exhibit strong discontinuities or irregularities. In the corotating frame, each ridge coincides with the maximum value of the Alfvén frequency along the radius, in agreement with \citetalias{dhouib_detecting_2022} and \citealp{barrault_exploring_2025} confirming that the magnetic field effectively acts as a low‑frequency filter, with modes below the Alfvén frequency becoming evanescent. As mentioned in \citet{barrault_exploring_2025}, this limit does not correspond to the mechanism presented in \citet{rui_gravity_2023} whereby the suppression mechanism is related to the radial component of the magnetic field.

\subsection{Comparison with current measurements}
\label{subsec:comparison_obs}
\begin{figure}[h]
    \centering
    \includegraphics[width=\columnwidth,trim=0cm 0.10cm 0cm 0.10cm, clip]{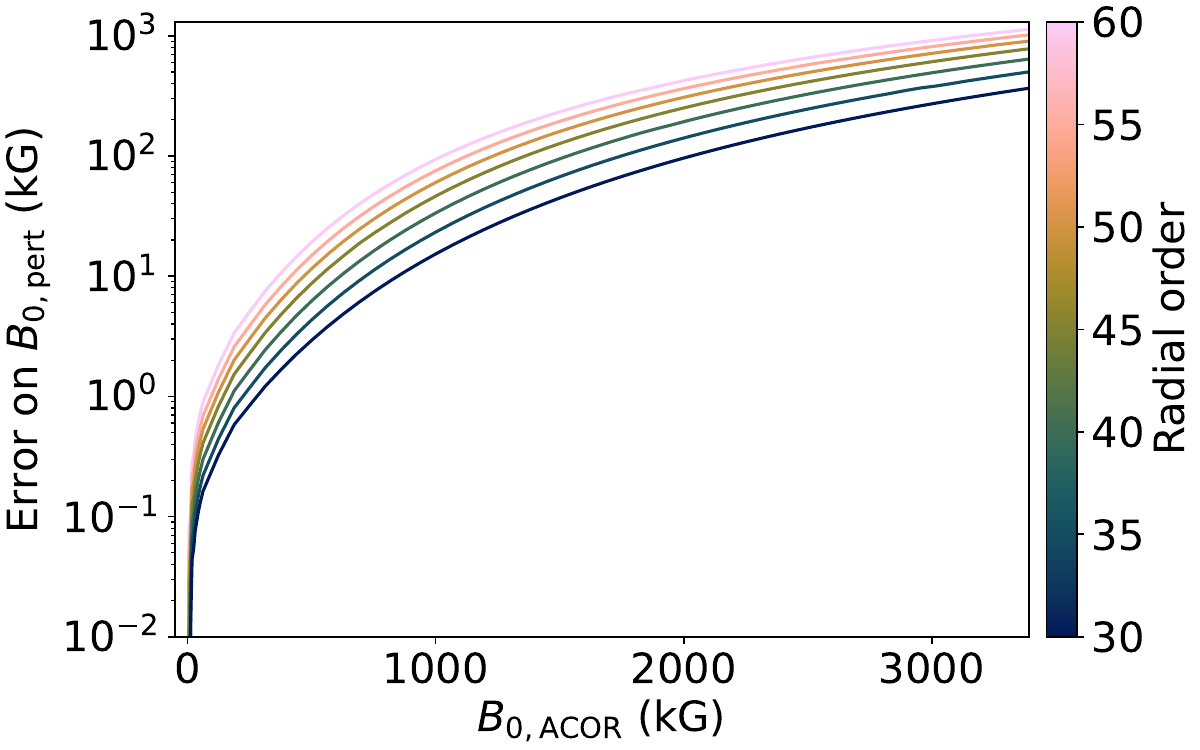}
    \caption{Estimation of the error on the intensity of the toroidal magnetic field by using a perturbative approach compared to the \acor approach. We used the same magnetic configuration as \citetalias{dhouib_detecting_2022} for the model \textit{ce4k}.}
    \label{fig:erroB0pert}
\end{figure}

Current measurements of internal magnetic field primarily concern red giant stars and yield detected values of $\langle B_r^2 \rangle^{1/2}$ in a range from a few tens of kilogauss to a few hundred kilogauss using different measurement techniques \citep{li_magnetic_2022,li_internal_2023,deheuvels_strong_2023,hatt_asteroseismic_2024,villate_seismic_2026}.
However, these measurements cannot be compared with the analysis presented in this study, which is restricted to toroidal magnetic fields.  A recent study \citet{takata_asteroseismic_2025} has reported the detection of a composite poloidal and toroidal magnetic field in a slowly rotating $\gamma$ Dor star. They found a lower bound of $92\pm 7$~kG for the root mean square of the toroidal component within 50 percent in radius from the center. In Fig. \ref{fig:erroB0pert}, we give an estimate of the error on the magnetic field strength from using a perturbative approach compared to the approach proposed here. We define the error on $B_0$ introduced by the perturbative approach, compared to \acor, as
\begin{equation}
    Err(B_{0,\mathrm{pert}}(\delta\nu))=|B_{0,\mathrm{\acor}}(\delta\nu)-B_{0,\mathrm{pert}}(\delta\nu)|,
\end{equation}
where $B_{0,\mathrm{\acor}}(\delta\nu)$ and $B_{0,\mathrm{pert}}(\delta\nu)$ correspond to the magnetic field strength, $B_0$, required to obtain a frequency shift, $\delta\nu$, for a specific mode computed with \acor and with the perturbative approach, respectively.
For the magnetic-field configuration studied here and for dipolar modes from $n=30$ to $n=60$, the error varies from approximately $0.2$~kG to $1.0$~kG for the same value of $B_{0,\mathrm{\acor}}\simeq 97$~kG. However, as explained in Appendix \ref{sec:termbyterm}, the topology used here is not compatible with the method used in their work so direct comparisons remain open to debate.

Finally, following Fig. \ref{fig:pert_Sc}, the frequency shift induced by a strong toroidal magnetic field may exhibit a $1/\omega_0^\alpha$ dependence, $\omega_0$ being the zeroth order angular frequency, with $\alpha$ increasing with the strength of the magnetic field. $\alpha$ could eventually reach values close to three, thereby mimicking the behaviour expected from the impact of a poloidal field in the perturbative framework. However, reaching such a regime requires a toroidal component significantly stronger than the radial component. This is because the perturbative approach remains valid when $B_\varphi/B_r \ll k_r/k_h\sim N/\omega\sim 10^2-10^3$ for the stellar model considered here. While such strong toroidal configurations are plausible and have been retrieved in simulations, their existence in real stars still needs to be observed.

\section{Conclusion}
\label{sec:conclusion}
In this study, we present the extension of the \acor oscillation code to include the effects of internally confined toroidal magnetic fields within a non-perturbative framework. The code was tested numerically through convergence tests on both the spectral and radial resolutions. The impact of this type of magnetic field on stellar oscillation modes was compared against the classical perturbative approach. In the slow-rotation and weak-field regime, the full calculations recover the perturbative predictions, thereby validating this new implementation of \acor. For higher magnetic field strengths, this new implementation has also been compared to the TARM implementation presented in \citetalias{dhouib_detecting_2022} and presents similar results. Deviations observed in stronger-field regimes highlight the importance of non-perturbative treatments when exploring stronger magnetic fields. Going beyond the TARM approximation enables asteroseismic studies in less stratified regions, allowing us to capture magnetic effects on the coupling between gravito-inertial modes in radiative regions and inertial modes in the convective core of intermediate-mass MS stars \citep{barrault_exploring_2025}. This opens the way for the seismic study of convective core magnetism, which will be investigated in a forthcoming work.

While this work relies on different assumptions, it provides a first step toward a more complete modeling of magnetic effects on stellar oscillations. We adopt spherical models with uniform rotation and a steady, axisymmetric, purely toroidal magnetic field confined to the interior, allowing us to isolate magnetic effects on mode dynamics. Relaxing these assumptions (by including structural distortions, poloidal or non-axisymmetric fields, or surface-reaching magnetic configurations) will require more advanced modeling and is left for future work.

This code is not tied to a specific stellar model. As long as it remains axisymmetric and within the framework of ideal MHD, it can be applied to different classes of pulsators and more complex toroidal magnetic fields. In addition, it enables the use of stellar models that are deformed by both rotation and magnetic fields. A natural next step is to include poloidal magnetic field configurations and extend the analysis to mixed modes in red giants. That would allow us to directly compare non-perturbative calculations to current asteroseismic constraints. In addition, parametric studies assessing the respective impacts of the stellar model, the rotation, and the magnetic field are necessary in order to disentangle the magnetic effects from other ones. Ultimately, these developments aim to establish robust magneto-asteroseismic diagnostics capable of constraining internal magnetic fields and improving our understanding of angular momentum transport in stellar interiors.

\begin{dataavailibility}
    The configuration file (.don file) listing the physical options and input data for Cesam2k20 calculations is given as \href{https://zenodo.org/records/21416531?token=eyJhbGciOiJIUzUxMiJ9.eyJpZCI6IjkzNTExNzg1LTJlY2QtNDNkMi1hMjRlLTE5MDgzMjI5YjJkMSIsImRhdGEiOnt9LCJyYW5kb20iOiI2M2YzYjc1ZTFkZDhjZTNmYzAyOWUwN2RjYzVmMjYyMSJ9.U43WWiJ04bdVwPKYXfDcNs_Vu7FbAi_Nc2Xr5ThTjDfwNh4zsgLH0RDVB8MyGVfzst5jLvh3131wA24aRHyQIg}{online supplementary material} for the model ce4k.
\end{dataavailibility}

\begin{acknowledgements}
    The authors thank Daniel Reese for insightful scientific and numerical discussions, as well as for computing and comparing modes in the nonmagnetic case with the TOP code. AF thanks Marie-Jo Goupil for her valuable discussions, careful corrections and guidance throughout the development of this implementation. We thank the referee for their thoughtful comments and constructive suggestions, which helped improve the quality and clarity of this manuscript. AF thanks Masao Takata for his valuable discussions and suggestions. This work was supported by the "Action Thématique de Physique Stellaire" (ATPS) of CNRS/INSU co-funded by CEA and CNES. AF, RMO, LM and LP acknowledge the {\it Action incitative de physique stellaire} (AIPS) of Paris Observatory's scientific council for their support. LB acknowledges the support of the Austrian Academy of Sciences through the Doctoral Fellowship Programme (DOC) of the Austrian Academy of Sciences 27648. LM acknowledges financial support from the French program "PROMETHEE" (Protostellar Magnetism: Heritage vs Evolution) managed by Agence Nationale de la Recherche (ANR) and from the ANR grant ANR-21-CE31-0018.
\end{acknowledgements}

\bibliographystyle{aa} 
\bibliography{references} 

\FloatBarrier
\begin{appendix}

\FloatBarrier

\section{Toroidal magnetic field models}
\label{sec:appendix_dhouib}

In this study as mentioned in \ref{subsec:models_mag}, we used two types of magnetic fields for the sake of simplicity and for comparison with previous studies \citep{mathis_probing_2021,bugnet_magnetic_2022} and in particular for the toroidal component in \citetalias{dhouib_detecting_2022}.
\subsection{Gaussian magnetic field}
\label{subsec:app_mag1}
\begin{figure}[h]
    \centering
    \includegraphics[trim=0cm 9cm 0cm 5cm, clip,scale=0.3]{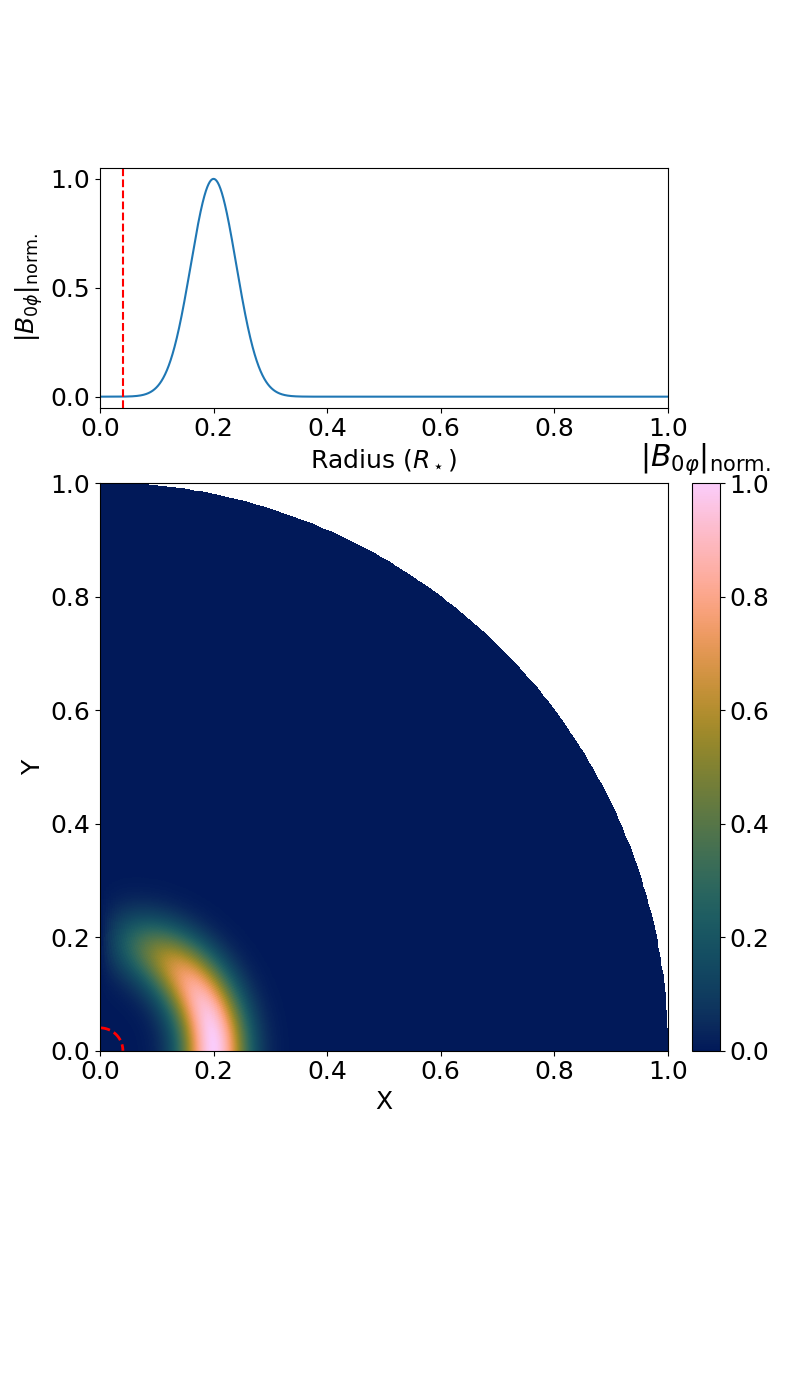}
    \caption{$|B_{o\varphi}|_{\rm{norm.}}$ is the absolute value of the toroidal magnetic field normalized by its maximum. Radial profile at the equator (\textit{upper panel}) and meridional plane (\textit{lower panel}) of $|B_{o\varphi}|_{\rm{norm.}}$. $B_{0\varphi}$ follows a Gaussian with parameters $r_0=0.2R_\star$ and $\sigma_0=0.04$. The red dotted line corresponds to the boundary convective core - radiative zone.}
    \label{fig:2Dmap_gaussian}
\end{figure}

The first one corresponds to a Gaussian radial profile (see Fig. \ref{fig:2Dmap_gaussian} for a representation): 
\begin{equation}
    B_{0,\varphi}(r,\theta) = B_0\mathrm{e}^{-\,(r-r_0)^2/(2\sigma_0)} \sin\theta,
\end{equation}

where $r_0$ and $\sigma_0$ are the parameters of the Gaussian and $B_0$ a normalization constant.

\subsection{Magnetic field following  \citetalias{dhouib_detecting_2022}}
\label{subsec:app_mag2}
The second type of magnetic field is the one presented in the work of \cite{duez_relaxed_2010} that provides a semi-analytic description of an axisymmetric mixed field with no rotation (we choose here to maintain the SI formulation as presented in the paper, as done in   \citetalias{dhouib_detecting_2022} where only the shape of the function is used):

\begin{equation}
    \vec{B_0}(r,\theta)=\dfrac{1}{r\sin(\theta)}\left(\vec{\nabla}\psi(r,\theta)\times\vec{e_\varphi}+\dfrac{\lambda \psi(r,\theta)}{R}\vec{e_\varphi}\right),
\end{equation}
\begin{equation}
    \psi(r,\theta)=\mu_0 \alpha \dfrac{A}{R}\sin^2(\theta) ,
\end{equation}
\begin{equation}
\begin{aligned}
A(r) &= -r \ \ \! \Bigg(
    j_1\!\left(\lambda \frac{r}{R}\right)
    \int_r^R y_1\!\left(\lambda \frac{x}{R}\right)\rho_0 x^3 \, dx \\
&\qquad\qquad
    + y_1\!\left(\lambda \frac{r}{R}\right)
    \int_0^r j_1\!\left(\lambda \frac{x}{R}\right)\rho_0 x^3 \, dx
\Bigg)
\end{aligned}
\end{equation}

with $\psi$ the magnetic stream function, $\mu_0$ the magnetic permeability, $\alpha$ a normalization constant fixed by the maximum strength of the magnetic field, $\lambda$ the eigenvalue to determine, $R$ the star radius, $j_1$ and $y_1$ the spherical Bessel functions of the second kind.

Even if this magnetic field has been shown to be unstable on timescales relevant to stellar evolution by \cite{kaufman_stability_2022}, it is used here for comparison purposes, in particular with the results of \citetalias{dhouib_detecting_2022} using the magnetic TAR extension. 

\begin{figure}[h]
    \centering
    \includegraphics[scale=0.27]{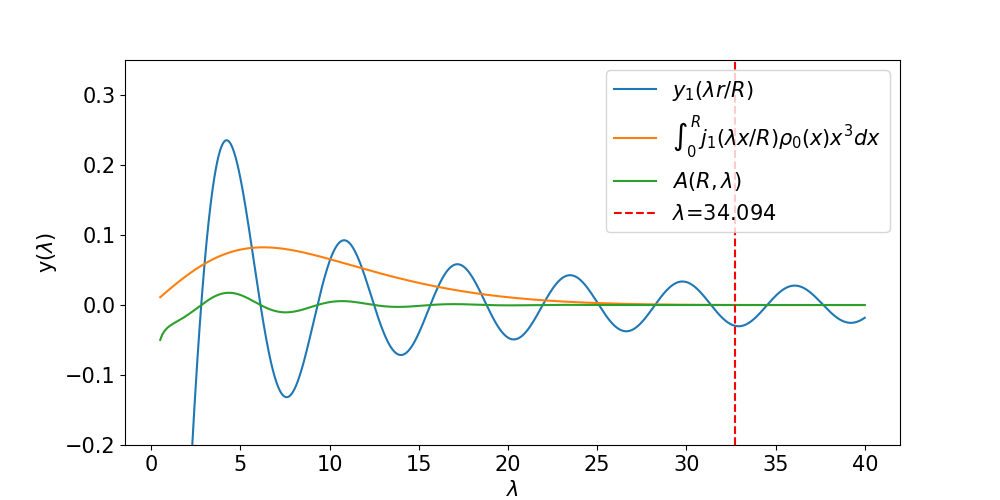}
    \caption{$A(R,\lambda)$ and its components for increasing values of $\lambda$. The red dotted vertical corresponds to the first positive value of $\lambda$ verifying $\vec{B_0}=\vec{0}$.}
    \label{fig:lambda_values}
\end{figure}

To find the value $\lambda$, we want that the $\vec{B_0}$ cancels out at the surface and not only the toroidal part of the magnetic field. This is obtained when the term $\int_0^rj_1(\lambda \dfrac{r}{R})\rho_0 x^3dx$ from $A(r,\lambda)$ cancels out at the surface. On the contrary, if we consider the cancellation of the term $y_1(\lambda \dfrac{r}{R})$ in $A(r,\lambda)$ as it was made in \cite{bugnet_magnetic_2022}, only the toroidal part is canceled out at the surface and not the poloidal. What we obtain corresponds to a shallower toroidal field. In Fig. \ref{fig:lambda_values} for the model \textit{ce4k}, we can see the values of these two terms $y_1(\lambda \dfrac{r}{R})$ and $\int_0^rj_1(\lambda \dfrac{r}{R})\rho_0 x^3dx$ at the surface as well as $A(r)$ for increasing $\lambda$. The first positive eigenvalue verifying $\vec{B_0}(R)=0$ is represented by the vertical line. 

\begin{figure}[h]
    \centering
    \includegraphics[trim=0cm 9cm 0cm 5cm, clip,scale=0.3]{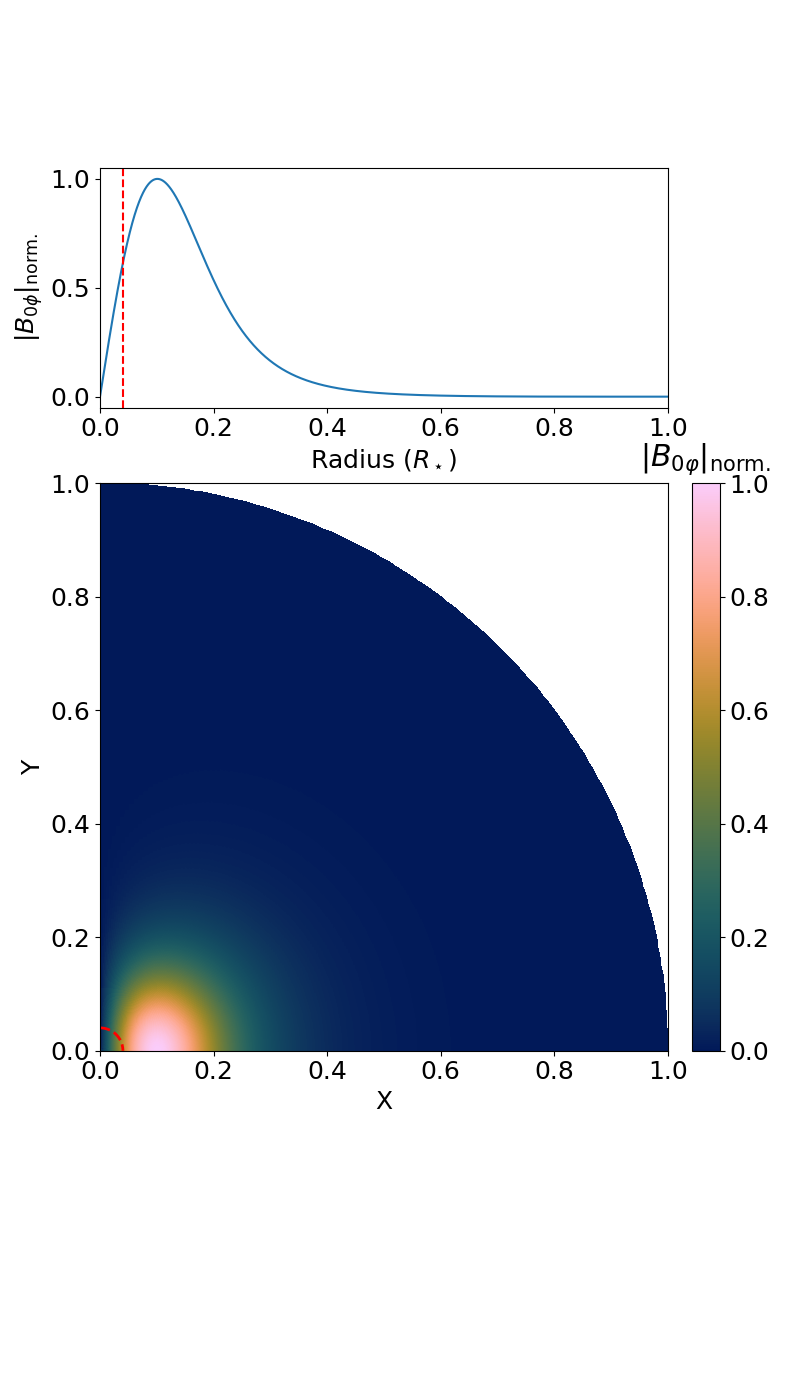}
    \caption{Same as Fig. \ref{fig:2Dmap_gaussian} but for a toroidal magnetic field following \citetalias{dhouib_detecting_2022}.}
    \label{fig:2Dmap_duez}
\end{figure}

In this study, we focus only on axisymmetric toroidal magnetic field. For that, we consider only the toroidal component of this semi-analytic description from \citep{duez_relaxed_2010} (see Fig. \ref{fig:2Dmap_duez} for a representation) following \citetalias{dhouib_detecting_2022}. 

\begin{equation}
    \vec{B}_{0\varphi}=B_0 \dfrac{b_\varphi(r)}{\rm{max}(|b_\varphi(r)|)}\sin(\theta)\vec{e_\varphi},
\end{equation}
with 
\begin{equation}
    b_\varphi(r)=\dfrac{\mu_0\alpha \lambda^2A(r)}{rR^2},
\end{equation}

We choose, as done in \cite{dhouib_detecting_2022}, to restrict the toroidal magnetic field to the radiative zone for comparison purposes.

\section{Stellar models}
\label{subsec:stellar_models}

The \acor code is independent of the specific stellar model used, but it does rely on several simplifying assumptions on the model in its current implementation. The structure is assumed to be axisymmetric and stationary on the oscillation time scale. The magnetic field is assumed to be toroidal and confined, and the set of equations is built in the MHD approach.

The stellar models used in this study are computed using the stellar evolution code Cesam2k20 (\citealp{morel_cesam_1997,morel_cesam_2008,Marques2013,Manchon2025}). As stated earlier, we focus on $\gamma$ Dor stars as precursors of red giants with detected core magnetic fields, offering a way to probe the evolution of internal magnetism.

To compute our Cesam2k20 models, we closely follow the choice of physical ingredients made in \citet{galoy_properties_2024}. We adopt the solar chemical composition of \citet{asplund_chemical_2009} with an initial helium and metal mass fraction of $Y_0 = 0.27$ and $Z_0 = 0.02$. In the radiative zones, the transport of chemical elements includes gravitational settling \citep{Michaud1993} and an ad hoc vertical turbulent mixing. Opacity tables adapted to the mixture are provided by the OPAL collaboration \citep{Rogers1992,Iglesias1996} and supplemented at low temperature by the Wichita opacity tables \citep{Ferguson2005}. The equation of state is also obtained from the OPAL2005 tables \citep{Rogers2002}. Nuclear reaction rates are sourced from the NACRE compilations \citep{angulo_compilation_1999}, except for the ${}^{14}{\rm N(p},\gamma){}^{15}{\rm O}$ reaction, for which LUNA's updated rates are used \citep{Broggini2018}. Convection is modeled with the mixing-length theory (MLT; \cite{bohm-vitense_uber_1958}), following the formulation of \citet{Henyey1965}, with a mixing-length parameter $\alpha=1.7$ and no overshoot. The atmosphere is reconstructed following a Hopf $T(\tau)$ relation \citep{Hubeny2014} and connected to the interior at a Rosseland optical depth of $\tau=20$.

The effect of rotation, while Cesam2k20 would allow it, is not accounted for in the stellar evolution modeling. Instead, we  tested the effect of a variety of ad hoc angular velocity profiles on the oscillation spectrum. Moreover, \acor is designed to compute oscillation spectra in non-spherical geometries, but the effect of centrifugal deformation in baroclinic stars, that Cesam2k20~can now model, is left to a future study.

The models used in this study are named {\it ceXk} where {\it ce} stands for Cesam2k20 and {\it Xk} corresponds to the number of grid points used to compute the modes --for instance the model \textit{ce4k} has been computed with approximately 4000 grid points. Their key characteristics are described in Table \ref{table:model_key}. The model \textit{ce4k} is used throughout this study unless specified otherwise.

\section{Analysis of the $1/\omega_0$ dependence of the splitting $S_c$}
\label{sec:termbyterm}

In this appendix, we compare our results obtained from the perturbative approach with the perturbation formulation given by Eq. (A63) of \citep{takata_asteroseismic_2025}. This expression predicts a $1/\omega_0$ dependence for the frequency shift induced by a toroidal magnetic field. However, for prograde dipolar modes, the magnetic frequency shift obtained with the perturbative approach in this study does not follow this scaling.

This discrepancy is expected. According to Eq.~(A63) of \citet{takata_asteroseismic_2025}, this integral cancels out due to the $\sin\theta$ dependence of this particular toroidal magnetic field. Moreover, the dominant contributions computed through the perturbative approach are not the same as in Eq.~(A38) of \citet{takata_asteroseismic_2025} where terms proportional to  $\left(\dfrac{\xi_h}{r}\right)^2$ dominate. When restricting the calculation to these terms, we find a negligible frequency shift.

\begin{figure}[h]
    \centering
    \includegraphics[trim=0cm 0cm 0cm 0cm, clip,scale=0.27]{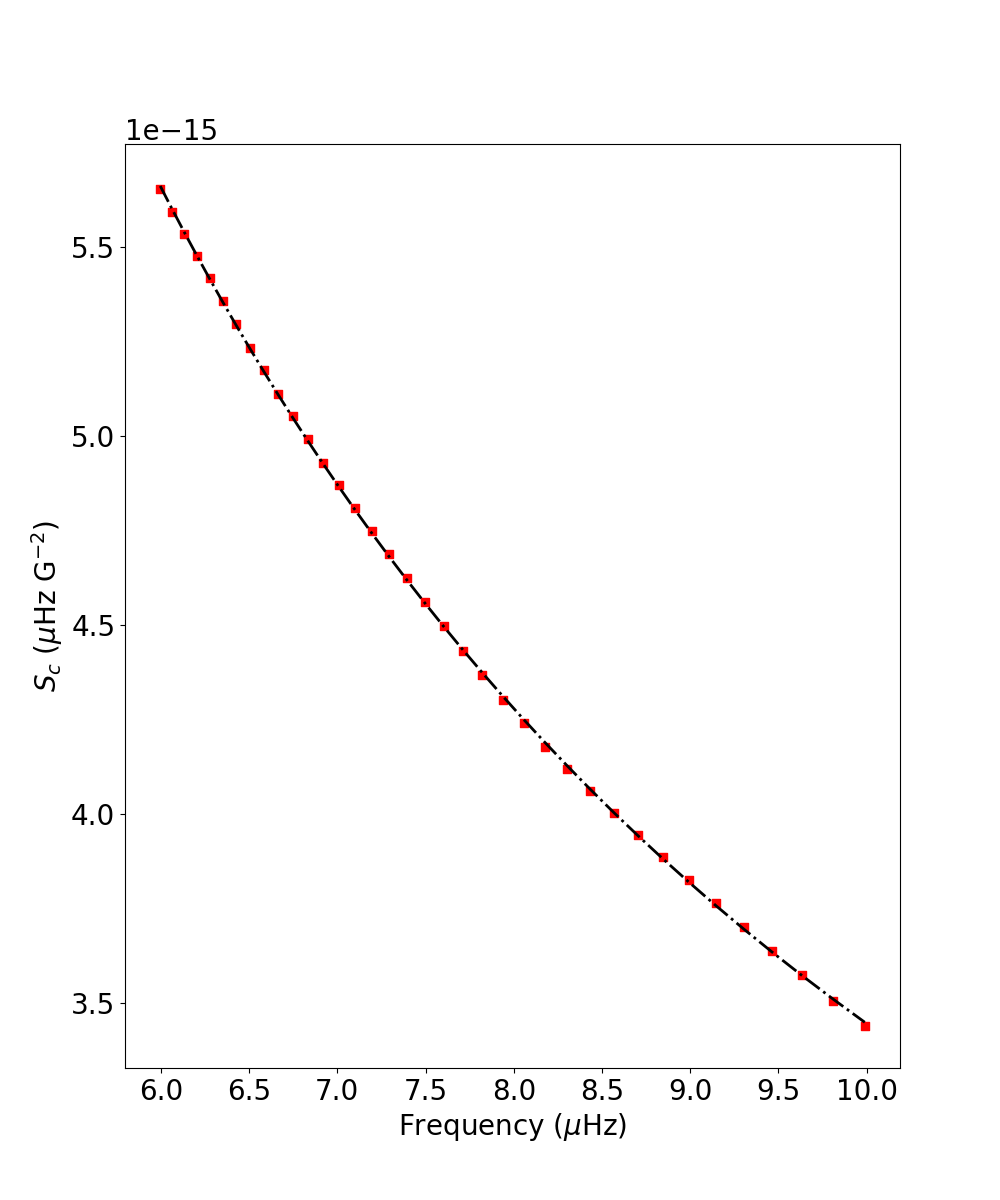}
    \caption{Splitting coefficient $S_c$ for quadrupolar prograde modes $(m=-1,l=2)$ along the frequency of the mode computed through the perturbative approach. A fit following $\delta\omega_{fit}=A\omega^\alpha$ is done with $A\approx 3.22\times 10^{-14}\mu{\rm Hz ~kG}^{-2}$ and $\alpha\approx -0.97$.}
    \label{fig:pert_Scl2}
\end{figure}

\begin{figure}[h]
    \centering
    \includegraphics[trim=0cm 0cm 0cm 0cm, clip,scale=0.27]{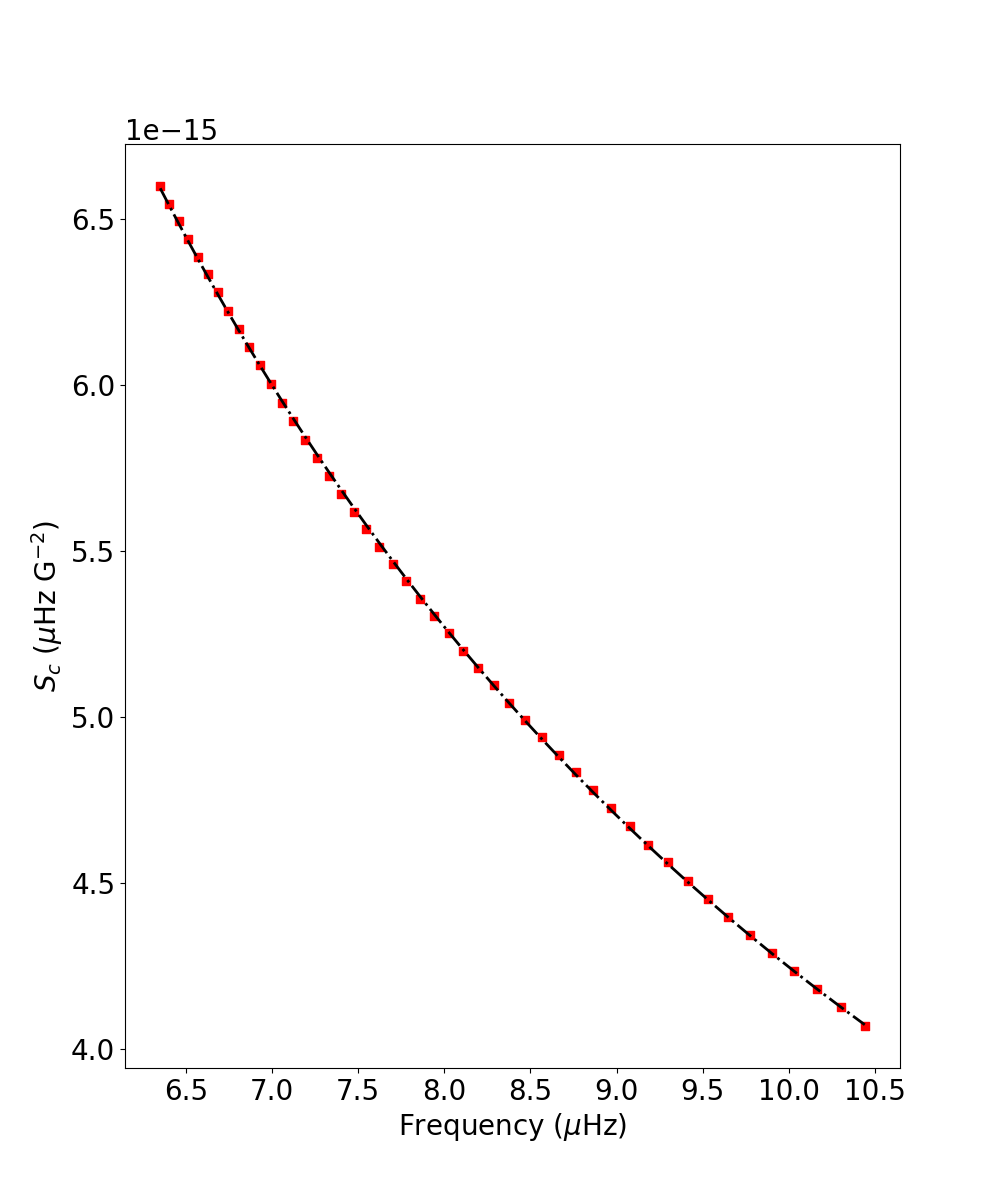}
    \caption{Splitting coefficient $S_c$ for octupolar prograde modes $(m=-1,l=3)$ along the frequency of the mode computed through the perturbative approach. A fit following $\delta\omega_{fit}=A\omega^\alpha$ is done with $A\approx3.95\times 10^{-14}\mu{\rm Hz ~kG}^{-2}$ and $\alpha\approx -0.97$.}
    \label{fig:pert_Scl3}
\end{figure}

For higher angular degree $\ell$, we recover a $1/\omega_0$ dependence in the splitting $S_c$ (see Fig. \ref{fig:pert_Scl2} and Fig. \ref{fig:pert_Scl3}).
For the mode $\ell=2,n=90$ (in the same frequency range as $\ell=1,n=60$), the integral of Eq.~(A38) in \citep{takata_asteroseismic_2025} does not cancel out. Consequently the $\left(\dfrac{\xi_h}{r}\right)^2$ terms dominate and it explains why we retrieve a $1/\omega_0$ dependence. 

These differences between our results and those of \citet{takata_asteroseismic_2025} most likely arise from the specific choice of toroidal magnetic field adopted in this study, whereas no particular field configuration is assumed in \citet{takata_asteroseismic_2025}.

\end{appendix}

\end{document}